\documentclass[12pt]{article}
\usepackage{amsfonts}
  \usepackage{a4wide}
  \usepackage{latexsym}
  \usepackage{epsf}
  \usepackage{amssymb}
  \usepackage{graphicx}
  \usepackage{amsmath,cite}
  \usepackage{slashed,epsfig}
  \usepackage{bbm}
  \usepackage{pstricks}
\newcommand{\ba}{\begin{eqnarray}}
\newcommand{\ea}{\end{eqnarray}}

\begin{document}
 %%%%%%%%%%%%%%%%%%%%%%%%%%%%%%%%%%%%%%%%%%%%%%%%%%%%%%%%%%%
\begin{titlepage}

\title
 {\bf Q7-branes and their coupling to IIB supergravity}
%{\bf Seven--brane coupling to IIB supergravity}

\author{
Eric Bergshoeff$^1$, Jelle Hartong$^1$ and Dmitri Sorokin$^{2,3}$
\\
\\
{\it \small $^1$ Centre for Theoretical Physics, University of Groningen},\\
{\it \small Nijenborgh 4, 9747 AG Groningen, The Netherlands}
\\
\\
{\it \small $^2$ INFN, Sezione di Padova}\\{\it \small ${\&}$
Dipartimento di Fisica ``Galileo Galilei", Universit\`{a} degli
Studi di Padova} \\
{\it\small via F. Marzolo 8, 35131 Padova, Italia}
\\
\\
{\it \small $^3$ Departamento de Fisica Te\'orica, Facultad de
Fisica,
Universidad de Valencia} \\
{\it\small C/Dr. Moliner, 50, 46100 Burjassot (Valencia), Espa\~na}}

\date{}
\maketitle

\begin{flushright}
\small\vskip -13truecm
\texttt{hep-th/yymmnnn}\\
UG-07-01\\
\end{flushright}

\thispagestyle{empty}

\vskip 12truecm

\begin{center}
{\bf Abstract}
\end{center}
We show how, by making use of a new basis of the IIB supergravity
axion-dilaton coset, $SL(2,\mathbb{R})/SO(2)$, 7-branes that
belong to different conjugacy classes of the duality group
$SL(2,\mathbb{R})$ naturally couple to IIB supergravity with
appropriate source terms characterized by an $SL(2,\mathbb{R})$
charge matrix Q. The conjugacy classes are determined by the value
of the determinant of Q. The $(p,q)$ 7-branes are the branes in
the conjugacy class ${\rm det}\,Q=0$. The 7-branes in the
conjugacy class ${\rm det}\,Q > 0$ are labelled by three numbers
$(p,q,r)$ which parameterize the matrix $Q$ and will be called
Q7-branes. We construct the full bosonic Wess--Zumino term for the
Q7-branes. In order to realize a gauge invariant coupling of the
Q7-brane to the gauge fields of IIB supergravity it is necessary
to introduce an $SL(2,\mathbb{R})$ doublet of two distinct Born--Infeld fields on the Q7-brane world-volume.

\end{titlepage}

\newpage
\tableofcontents

\newpage
\setcounter{equation}0
\section{Introduction}

In \cite{Bergshoeff:2006jj} 7-brane solutions of IIB
supergravity have been investigated with an emphasis on their
supersymmetry properties. One of the results of that paper was
the observation that the construction of globally well-defined supersymmetric 7-brane
solutions from a 10-dimensional point of view can be realized
provided one introduces a new type of 7-brane.

It has been proposed in \cite{Meessen:1998qm} that 7-branes are
described by a triplet of charges here denoted by $(p,q,r)$. This is
in contrast with the common statement that in type IIB superstring
theory there are only $(p,q)$ 7-branes. The main argument of
\cite{Meessen:1998qm} in support of this idea was that the RR 8-form
transforms as part of a triplet of 8-forms under $SL(2,\mathbb{R})$~\footnote{It is understood that the $SL(2,\mathbb{R})$ duality group is quantized and given by $SL(2,\mathbb{Z})$. Whenever we speak of $SL(2,\mathbb{R})$ we mean to imply that the result under discussion does not depend on the $SL(2,\mathbb{Z})$ charge quantization.}.
This observation was also made in \cite{Dall'Agata:1998va}. The
$(p,q,r)$ charges naturally parameterize an $SL(2,\mathbb{R})$ algebra valued
matrix $Q$
\begin{equation}\label{Q}
    Q=\left(\begin{array}{cc}
        r/2 & p \\
        -q & -r/2
        \end{array}
        \right)\,.
\end{equation}
So an arbitrary $SL(2,\mathbb{R})$ transformation can be written as
$e^Q$. In \cite{Bergshoeff:2002mb} it was shown that at least locally one can write
down three families of 7-brane solutions parameterized by the value
of
$\text{det}\,Q$. The three families depend on whether
$\text{det}\,Q<0$, $\text{det}\,Q=0$ or $\text{det}\,Q>0$. The D7-brane corresponds to $p=1$ and $q=r=0$, i.e. it has $\text{det}\,Q=0$.

What motivated the research which led to \cite{Bergshoeff:2006jj}
was to reconcile these ideas with the globally well-defined
supersymmetric F-theory solutions. This reconciliation has been
successful and has lead to the conclusion that for $\text{det}\,Q<0$
there are no well-defined 7-brane solutions while the
 branes with $\text{det}\,Q>0$ play an important role in the construction of F-theory $(p,q)$
7-brane solutions from a 10-dimensional point of view. In \cite{Bergshoeff:2006jj} it was further shown that the well-known F-theory 7-brane solutions form a subset of a much larger set of globally well-defined and supersymmetric 7-brane configurations. This larger set of solutions contains 7-branes with $\text{det}\,Q>0$. In this paper we study in
more detail the properties of the 7-branes with $\text{det}\,Q>0$.
They will be referred to as `Q7-branes'.

In the analysis of
\cite{Bergshoeff:2006jj} a so-called pseudo-action (which includes
7-brane source terms) was used to describe the
$SL(2,\mathbb{R})$ invariant coupling of the 7-branes to the
axion $\chi$ and the dilaton $\phi$ of IIB supergravity combined
into a complex field $\tau=\chi+ie^{-\phi}$. One of the features of
the pseudo-action of
\cite{Bergshoeff:2006jj} is that the 7-brane source term does
not contain a Wess--Zumino term describing the minimal electric
coupling of the 7-brane to an 8-form field. The reason one can leave
out such a term is because 8-forms are dual to the scalars and a
source term for the scalars is provided by the Nambu--Goto term. If
one then assumes holomorphicity of $\tau$ the Nambu--Goto source
acts for both the real and imaginary part of $\tau$.

Part of the motivation for this article has been to improve on this
situation by describing 7-brane coupling to an 8-form field via the
Wess--Zumino term. We will explain how upon the duality
transformation which eliminates the 8-form field, the information
about the magnetic coupling of the 7-brane to the axion-dilaton is
encoded in a non-locality associated with the presence of a Dirac
brane
\cite{Dirac,Deser:1997mz,Medina:1997fn,Bandos:1997gd} stemmed from
the 7-brane. The Dirac branes are associated with the branch cut
properties of the holomorphic function $\tau$ describing the
corresponding supersymmetric 7-brane solution.

Once we have identified the Q7-branes in the static limit with zero Born--Infeld (BI) vector fields we proceed to study
the world-volume theory of the Q7-branes. It will be shown that the gauge invariant coupling
of the Q7-brane to the IIB supergravity 8-, 6-, 4- and 2-forms, described
by the Wess--Zumino term, requires the introduction of two distinct
BI fields. The two distinct BI vectors transform as a linear doublet under $SL(2,\mathbb{R})$. We further argue that both of these BI fields are propagating on the world-volume
by constructing the Dirac--Born--Infeld action up to second order in the BI vectors.

The paper is organized as follows. In Section \ref{sec:D7coupling}
we discuss in detail how a D7-brane couples electrically to the RR
8-form and magnetically to the RR axion via the presence of a Dirac
8-brane. This sets the stage for the discussion of the coupling of
the Q7-branes in Section \ref{sec:Q7coupling}. To describe the Q7-brane coupling new
coordinates, denoted by $T$ and $\chi'$, for the axion-dilaton
coset manifold $SL(2,\mathbb{R})/SO(2)$ are introduced. In Section
\ref{sec:Diracstrings} we discuss the relation between the Dirac
brane stemming from the Q7-brane and the monodromy of the
axion-dilaton field $\tau$ measured when going around a Q7-brane.
In Section
\ref{sec:conjugacyIIB} the relation between the axion-dilaton fields
$(\chi,\phi)$ and the fields $(\chi',T)$ is derived. In Section
\ref{wvaction} we describe how the Q7--branes couple to the 8-, 6-, 4-
and 2-forms and to the axion--dilaton and in Section
\ref{sec:discussion} we end up with a summary
of the results and with a discussion of some issues regarding the
nature of the Q7--branes.

\setcounter{equation}0
\section{D7-brane coupling}\label{sec:D7coupling}
To illustrate the method of how to couple magnetically charged
branes to IIB supergravity it is instructive to first consider the
example of the conventional Dirichlet 7-brane. In this particular
case the coupling to the corresponding sector of IIB supergravity
follows the classical rules of how to describe magnetically charged
particles \`a la the Dirac monopole \cite{Dirac}. This has been
extended to cases of various magnetically charged brane sources in
\cite{Deser:1997mz,Medina:1997fn,Bandos:1997gd}.

The D7-brane world-volume action in the Einstein frame (which is
appropriate for our purposes) has the following form
\cite{Cederwall:1996ri}
\begin{equation}\label{DBI}
{{S}}= -\int_{{{\mathcal M}_8}}\,d^{8}\xi\,e^{ {\phi}}\sqrt{-\det
(g_{\mu\nu}+e^{-{1\over 2} {\phi}}{\mathcal F}_{\mu\nu})}-
\int_{{{\mathcal M}_8}}\,{\mathcal{C}}\wedge e^{{\mathcal F}_2}
\,,
\end{equation}
where ${{\mathcal M}_8}$ is the 8-dimensional world-volume
parameterized by $\xi^{\mu}$, with $\mu=0,1,\ldots,7$ and where
$\phi(\hat x(\xi))$ and $g_{\mu\nu}$ are the pullbacks of the
dilaton and the target space metric onto the world-volume,
respectively. In the Wess--Zumino term $\mathcal{C}$ denotes the
formal sum of the pullbacks of the (duality related) RR potentials
$C_{r}(\hat x(\xi))$ $(r=0,2,4,6,8)$ and ${\mathcal
F}_{2}=d{\mathcal A}_{1}+B_{2}$ is the field strength of the
world-volume Born--Infeld gauge field ${\mathcal A}_{1}$ extended
with the pullback of the NSNS 2-form $B_{2}$.

In what follows, we shall discuss the coupling of the brane action in which the
Born--Infeld field has been put to zero, i.e. ${\mathcal F}_{2}=0$. In other words the
D7-brane will be coupled only to the RR 8-form potential
$C_{8}$ dual to the axion field $C_{0}=:\chi$. The reason is
that for the Q7-branes whose charges are not in the same
$SL(2,\mathbb{R})$ conjugacy class as those of the D7-brane, the
complete Born--Infeld part of the action is unknown. We shall
present the Q7-brane Wess--Zumino term with two Born--Infeld fields
in Section \ref{wvaction}.

In the absence of the Born--Infeld field the action (\ref{DBI})
reduces to
\begin{equation}\label{NG+WZ}
{{S}}= -\int_{{{\mathcal M}_8}}\,d^{8}\xi\,e^{
{\phi}}\sqrt{-g_{(8)}}- \int_{{{\mathcal M}_8}}\, {C}_{8} \,,
\end{equation}
where $\sqrt{-g_{(8)}}$ is used to denote $\sqrt{-\det
 ~g_{\mu\nu}}$.

We would like to couple this action to a corresponding part of IIB
supergravity action, assuming that the NSNS and RR 2-form
fields are zero (which otherwise would produce sources for a
world-volume gauge field).

Since in (\ref{NG+WZ}) instead of the axion we have its dual 8-form
potential $C_8$, we are not allowed to take the IIB supergravity
action with the axion-dilaton sector in the conventional form
\begin{equation}\label{ad}
S=\int\,d^{10}x\,\sqrt{-g_{(10)}}\,\left(R-{1\over
2}\,{\partial_m}\phi\,{\partial^m}\phi-{1\over
2}\,e^{2\phi}\,\partial_m \chi\,\partial^m \chi\right)\,,
\end{equation}
where $C_{8}$ does not appear.

The issue is that we should work, both in the bulk as well as in the
source term, with either $C_{8}$ or $\chi$. It seems natural to
electrically couple the D7-brane to $C_{8}$, but most of the
7-brane calculations are done in a formulation in which the
7-brane magnetically couples to the axion.
Therefore, the best strategy is not to work with either of the two
formulations but instead take as the starting point the following
first order action which, as we will see, interpolates between the
two formulations:

\begin{eqnarray}\label{ad+D7}
S=\int\,d^{10}x\,\sqrt{-g_{(10)}}\,\left(R-{1\over
2}\,{\partial_m}\phi\,{\partial^m}\phi-{1\over 2}\,e^{2\phi}\,F_m
\,F^m + {1\over{8!\sqrt{-g_{(10)}}}}\epsilon^{mn_1\cdots
n_9}\,F_m\,\partial_{n_1}C_{n_2\cdots
n_9}\right)\nonumber\\
\\
-\int_{\mathcal M_8}\,d^{8}\xi\,e^{ {\phi}}\sqrt{-g_{(8)}}-
\int_{\mathcal M_8}\, {C}_{8} \,.\hspace{150pt} \nonumber
\end{eqnarray}
Here $F_m$ is now an auxiliary vector field which replaces the axion
derivative and $C_{8}$ is its dual. Without the source term, Eq.
\eqref{ad+D7} establishes the standard duality relation between
the fields $\chi$ and $C_{8}$.

To perform the duality transform in the presence of the brane
source, we should rewrite the second line of Eq. \eqref{ad+D7}
as a 10D bulk integral. To this end we introduce the D7-brane
current 8-form with the delta function having support on the
D7-brane world-volume
\begin{equation}\label{j}
J^{m_1\cdots m_8}={1\over{\sqrt{-g_{(10)}}}}\int_{{\mathcal M}_8} d
\hat x^{m_1}\wedge\cdots\wedge d\hat x^{m_8} \delta(x-{\hat
x}(\xi))\,.
\end{equation}
The 2-form dual of the current is the closed form
\begin{equation}\label{sarj}
({~}^*J)_{m_1m_2}={1\over{ 8!}} \epsilon_{m_1m_2n_1\cdots
n_8}\int_{{\cal M}_8} d{\hat x}^{n_1}\wedge\cdots\wedge d{\hat
x}^{n_8}\delta(x-{\hat x}(\xi))\,,\qquad d{~}^*J_{8} =0\,.
\end{equation}
As such, at least locally, we can represent ${~}^*J_{8}$ as the
differential of a 1-form which we shall call ${~}^*G_{9}$, namely
\begin{equation}\label{sarj1}
{~}^*J_{8}=d{~}^*G_{9} \quad \Rightarrow \quad
({~}^*G)_m={1\over{9!}} \epsilon_{mn_1\cdots n_9}\int_{{\cal M}_9}
d{\hat x}^{n_1}\wedge\cdots\wedge d{\hat x}^{n_9}\delta(x-{\hat
x}(y))\,
\end{equation}
and
\begin{equation}\label{sarj2}
G^{n_1\cdots n_9}={1\over{\sqrt{ -g_{10}}}}\, \int_{{\cal M}_9}
d{\hat x}^{n_1}\wedge\cdots\wedge d{\hat x}^{n_9}\delta(x-{\hat
x}(y))\,.
\end{equation}
In the last equation the delta function has the support on a
9-dimensional surface ${\cal M}_9$, parameterized by coordinates
$y$, whose boundary is the world-volume ${\mathcal M}_{8}$ of the
D7-brane. The 9-dimensional surface is associated with the
world-volume of a Dirac 8-brane, which is a brane generalization
of the Dirac string stemming from a monopole. In the present case
we have a Dirac 8-brane stemming from the D7-brane. It is by means
of the Dirac 8-brane that the D7-brane will magnetically couple to
the axion field strength $F_{1}=d\chi$ as we shall see.

With the help of the dual current \eqref{sarj} the second line of
Eq. \eqref{ad+D7} can be rewritten as a 10D integral as follows
\begin{eqnarray}\label{ad+D7+10}
S=\int_{\mathcal M_{10}}\,d^{10}x\,\sqrt{-g_{(10)}}\,\left(R-{1\over
2}\,{\partial_m}\phi\,{\partial^m}\phi-{1\over 2}\,e^{2\phi}\,F_m
\,F^m + {1\over{8!\sqrt{-g_{(10)}}}}\epsilon^{mn_1\cdots
n_9}\,F_m\,\partial_{n_1}C_{n_2\cdots
n_9}\right)\nonumber\\
\\
-\int_{\mathcal M_{10}}\,d^{10}x\int_{\mathcal
M_8}\,d^8\xi\,\delta(x-\hat x(\xi))\,e^{ {\phi}}\sqrt{-g_{(8)}}-
\int_{\mathcal M_{10}}\, {C}_{8}\wedge {~}^*J_{8} \,, \nonumber
\end{eqnarray}
where we use the convention that $\epsilon_{01\ldots
9}=-\epsilon^{01\ldots 9}=1$. We will use this first order action to
derive expressions for the bulk plus source terms with $\chi$ or
$C_{8}$ only.

We now wish to first eliminate from Eq. \eqref{ad+D7+10} the
field $C_{8}$ by solving the equations of motion of $F_{1}$ and
$C_{8}$. The variation of \eqref{ad+D7+10} with respect to
$F_{1}$ gives the duality condition
\begin{equation}\label{dual81}
F_{1}=e^{-2\phi}\,{}^*dC_{8}=:e^{-2\phi}\,{}^*F_{9}\,.
\end{equation}
The variation with respect to $C_{8}$ gives
\begin{equation}\label{eq8s0}
d\,F_{1}=-{}^*J_{8}\,.
\end{equation}
Thus $F_{1}$ is not a closed form. However, recalling that the dual
current $*J_{8}$ is the exact form \eqref{sarj}, we have
\begin{equation}\label{eq8s}
d\,F_{1}=-{}^*J_{8}=-d{}^*G_{9}\quad\Rightarrow\quad
d(F_{1}+{}^*G_{9})=0\,.
\end{equation}
Hence, at least locally,
\begin{equation}\label{1dchis}
F_{1}+{}^*G_{9}=d\chi\,, \quad \Rightarrow \quad
F_{1}=d\chi-{}^*G_{9}\,,
\end{equation}
where $\chi$ is the axion.

We can now eliminate the field $C_{8}$ from the action. To do this
we note that up to a total derivative (which we shall skip) the last
term in
\eqref{ad+D7+10} can be rewritten as
\begin{equation}\label{last}
-\int_{\mathcal M_{10}}\, {C}_{8}\wedge {~}^*J_{8}=\int_{{\mathcal
M}_{10}}\left(dC_{8}\wedge {}^*G_{9}+{\rm total~~derivative}\right)\,.
\end{equation}
Let us now substitute in the action (\ref{ad+D7+10}) the expression
\eqref{1dchis} for the auxiliary field $F_{1}$. Then the term
\eqref{last} and the last term in the first line of \eqref{ad+D7+10}
cancel each other (up to a total derivative). As a result we arrive
at the action
\begin{eqnarray}\label{ad+D7-WZ}
S_{\text{I}}=\int_{\mathcal
M_{10}}\,d^{10}x\,\sqrt{-g_{(10)}}\,\left(R-{1\over
2}\,{\partial_m}\phi\,{\partial^m}\phi-{1\over 2}\,e^{2\phi}\,F_m
\,F^m \right) -\int_{\mathcal M_8}\,d^8 \xi\,e^{
{\phi}}\sqrt{-g_{(8)}}\,,
\end{eqnarray}
where $F_{1}=dx^mF_m$ is defined in (\ref{1dchis}). We see that the
minimal coupling of the D7-brane to $C_{8}$ disappears and its role
is taken upon by the non-local terms in the axion field strength
$F_{1}$ due to the contribution of the Dirac 8-brane.

Alternatively, we can eliminate the axion in favor of the RR
field $C_{8}$. To this end we apply the equation of motion
corresponding to the auxiliary field $F_m$ leading to the duality
relation \eqref{dual81}. We use this relation to eliminate $F_m$ and
this leads to the following action
\begin{eqnarray}\label{actionII}
S_{\text{II}}&=&\int\,d^{10}x\,\sqrt{-g_{(10)}}\,\left(R-{1\over
2}\,{\partial_m}\phi\,{\partial^m}\phi-{1\over {2\cdot
9!}}\,e^{-2\phi}\,F_{m_1\cdots m_9} \,F^{m_1\cdots m_9} \right)\cr
&&-\int_{{{\mathcal M}_8}}\,d^{8}\xi\,e^{ {\phi}}\sqrt{-g_{(8)}}-
\int_{{{\mathcal M}_8}}\, {C}_{8}\,,
\end{eqnarray}
where $F_9=dC_8$. Since the actions $S_{\text{I}}$ and
$S_{\text{II}}$ follow from the same action S given in
\eqref{ad+D7+10} they are classically equivalent.

\setcounter{equation}0
\section{Q7-brane coupling}\label{sec:Q7coupling}
In the generic case of a $(p,q,r)$ 7-brane (referred to as the
Q7-brane) the Nambu--Goto and the Wess--Zumino part of the 7-brane
action has the following form \cite{Bergshoeff:2006ic}
\begin{equation}\label{NG+WZG}
{{S}}=-\int_{{{\mathcal M}_8}}\,d^{8}\xi\,\,T \,\sqrt{-g_{(8)}}-
\int_{{{\mathcal M}_8}}\, q_{\alpha\beta}\,A^{\alpha\beta}_{8} \,,
\end{equation}
where $T$ given by
\begin{equation}\label{tension}
  T=q_{\alpha\beta}\,V^{\alpha}_-V^{\beta}_+
  =\frac{1}{\text{Im}\,\tau}\left(p+r\,\text{Re}\,\tau +q\vert\tau\vert^2\right)
\end{equation}
is the so-called `tension scalar' (for the definition of
$V_{\pm}^{\alpha}$ and related quantities we refer the reader to
Appendix \ref{coset}). In Subsection \ref{cosys} it is shown that for $q\neq 0$ we have $\text{sign}(q)\,T\ge 2\sqrt{\text{det}\,Q}$ so that the sign
of the tension is determined by the sign of the parameter $q$~\footnote{The tension $T$ of a 7-brane is negative when $q<0$. The negative tension Q7-branes play a similar role as orientifold O8-planes play in the case of the D8-brane solutions \cite{Bergshoeff:2001pv}; they are used to make the supergravity solutions globally well-defined. Orientifold 7-planes only show up when the axion-dilaton field $\tau$ is constant \cite{Sen:1996vd}. For the case of non-constant $\tau$ we need negative tension $Q7$-branes \cite{Bergshoeff:2006jj}.\label{negativetension}}. The 8-forms
$A^{\alpha\beta}_{8}$ form an $SU(1,1)$ triplet among
which only two 8-forms are independent due to the following constraint on
their field strengths \cite{Dall'Agata:1998va} (see also Appendix
\ref{sec:IIB}, Eq. \eqref{definitionGtilde})
\begin{equation}\label{constraintonA}
V^{\alpha}_-V^{\beta}_+\,F_{9\,\alpha\beta}=0\,.
\end{equation}
The charge tensor $q_{\alpha\beta}$ transforms in the adjoint of
$SU(1,1)$. The $SU(1,1)$ indices $\alpha, \beta$ are raised and
lowered with the 2-dimensional epsilon symbol
$\epsilon_{12}=\epsilon^{12}=1$, i.e. $q^{\alpha}_{\;\;\;\beta}=\epsilon^{\alpha\gamma}q_{\gamma\beta}$
for raising and
$q_{\alpha}^{\;\;\;\beta}=q^{\gamma\beta}\epsilon_{\gamma\alpha}$
for lowering the indices. The action \eqref{NG+WZG} is
$SU(1,1)$ invariant provided we also transform
the charges
$q_{\alpha\beta}$.
%We define
%\begin{eqnarray}
% & p=\frac{1}{4}(q^{11}+q^{22}-2q^{12})\label{p}\,, \\
% & q=-\frac{1}{4}(q^{11}+q^{22}+2q^{12})\label{q}\,, \\
% & r=\frac{i}{2}(q^{11}-q^{22})\label{r}\,.
%\end{eqnarray}
We introduce the $SL(2,\mathbb{R})$ algebra valued charge matrix $Q$
defined by
\begin{equation}\label{matrixQ}
    Q=-{i\over 2}S^{-1}q\epsilon S=\left(
    \begin{array}{cc}
     r/2 & p \\
     -q & -r/2
     \end{array}
    \right)\,, \qquad S=\left(\begin{array}{cc}
                            -i & 1 \\
                            i & 1
                            \end{array}\right)\,,
\end{equation}
where $q\epsilon$ is a matrix whose components are given by
$(q\epsilon)^{\alpha}_{\;\;\;\beta}=q^{\alpha\gamma}\epsilon_{\gamma\beta}$.
The matrix $Q$ was mentioned in the Introduction, Eq.
\eqref{Q}. The matrix $S$ establishes the isomorphism between
$SL(2,\mathbb{R})$ and $SU(1,1)$.

Each value of ${\rm det}\,Q $
forms an
$SL(2,\mathbb{R})$ conjugacy class. The conjugacy classes of $SL(2,\mathbb{R})$ are characterized by the
value of the trace of the $SL(2,\mathbb{R})$ matrix
\begin{equation}\label{eQ}
    e^Q=\cos(\sqrt{\text{det}\,Q})\mathbbm{1}+\frac{\sin(\sqrt{\text{det}\,Q})}{\sqrt{\text{det}\,Q}}Q\,.
\end{equation}
The families of conjugacy classes are formed by
\begin{equation}
{\rm Tr}\,e^Q=2\cos(\sqrt{\text{det}\,Q})\,\left\{
\begin{array}{c}
=2\\
>2\\
<2
\end{array}\right.\quad\text{or by}\quad \text{det}\,Q\,\left\{\begin{array}{c}
=0\\
<0\\
>0
\end{array}\right.
\end{equation}
When $\det\,Q=0$ we are in the conjugacy class to which the $(p,q)$ 7-branes
belong. The D7-brane corresponds to
$p=1$ and $q=r=0$.

\subsection{Q7-branes in the $(T,\chi')$ basis}\label{c8chi'}

In the case of the D7-brane the RR 8-form is dual to the RR scalar
and the RR scalar does not appear in the Nambu--Goto part, so that
in some sense the degrees of freedom described by the Nambu--Goto
and Wess--Zumino part are `orthogonal'. In the case of the Q7-brane
action (\ref{NG+WZG}) the Nambu--Goto term contains a non-linear
combination $T = q_{\alpha\beta}\,V^{\alpha}_-V^{\beta}_+$ of the
dilaton and axion fields. The question is whether we again have that
the scalar which is dual to $q_{\alpha\beta}\,A^{\alpha\beta}_{8}$
and the scalar $T$ are independent or `orthogonal'. Therefore one
must know to which axion-dilaton function the field
$q_{\alpha\beta}\,A^{\alpha\beta}_{8}$ is dual to and whether or not
this composite scalar field appears as part of $T$ in the
Nambu--Goto term. If the latter were the case, this would
significantly complicate the dualization procedure, since it would
then require the explicit use of the PST formalism
\cite{Pasti:1995tn} and the proof that the coupling of the 7-brane
to the duality symmetric type IIB supergravity
\cite{Dall'Agata:1997ju,Dall'Agata:1998va} obeys the PST symmetries,
as {\it e.g.} in the case of the M5-brane coupled to 11D
supergravity \cite{Bandos:1997gd}.

As it turns out the tension scalar $T$ in the Nambu--Goto term is
completely orthogonal (on the mass shell) to the scalar field which
is the dual of $q_{\alpha\beta}\,A^{\alpha\beta}_{8}$, to be called
$\chi'$, in the sense that both have diagonal kinetic terms, see
below. This diagonalization occurs due to the fact that the
combination $V^{(\alpha}_-V^{\beta)}_+$, which appears in the
definition of the scalar $T$, is orthogonal to $F_{9}^{\alpha\beta}$
by virtue of the constraint (\ref{constraintonA}).

In order to rewrite the axion-dilaton kinetic terms in terms of $T$
and $q_{\alpha\beta}A_{8}^{\alpha\beta}$ we first observe that the
derivative of $T$ can be written as
\begin{equation}
    dT=q_{\alpha\beta}V_{+}^{\alpha}V_{+}^{\beta}\bar
    P+q_{\alpha\beta}V_{-}^{\alpha}V_{-}^{\beta}P\,,
\end{equation}
where $P$ is defined in Eq. \eqref{defP}. At the same time,
the duality relation can be written as
\begin{equation}\label{duality}
    F_{9}^{\alpha\beta}=-i*\left(V_{-}^{\alpha}V_{-}^{\beta}P
    -V_{+}^{\alpha}V_{+}^{\beta}\bar P\right)\,,
\end{equation}
see Eq. \eqref{F9duality}. It is now straightforward to show
that we can write the scalar field kinetic terms of the IIB
supergravity action as
\begin{equation}\label{Tphi}
{\partial_m}\phi\,{\partial^m}\phi+e^{2\phi}\,\partial_m
\chi\,\partial^m \chi \, =\,
\frac{1}{T^2-4\det\,Q}\,\left({\partial_m}T{\partial^m}T+\frac{1}{9!}\,
q_{\alpha\beta}F_{m_1\cdots
m_9}^{\alpha\beta}q_{\gamma\delta}F^{\gamma\delta\,\,m_1\cdots
m_9}\right)\,,
\end{equation}
where $F^{\alpha\beta}_9$ has been defined in (\ref{duality}). With this observation the dualization procedure of the coupled
IIB supergravity -- Q7-brane system proceeds analogously to the case of the
D7-brane, but with the fields $T$ and
$q_{\alpha\beta}\,A^{\alpha\beta}_{8}$ instead of $\phi$ and
$C_{8}$. We start with the following first-order action which is
similar to (\ref{ad+D7+10})
\begin{eqnarray}\label{ad+7+10}
S=\int_{\mathcal
M_{10}}\,d^{10}x\,\sqrt{-g_{(10)}}\,\left(R-{{{\partial_m}T{\partial^m}T}\over
{2\left(T^2-4{\rm det}~Q\right)}}\,-{1\over 2}\left(T^2-4{\rm
det}~Q\right)\,F_mF^m \right.\hspace{30pt}
\nonumber\\
\\
\left.+ {1\over{8!\sqrt{-g_{10}}}}\epsilon^{mn_1\cdots
n_9}\,F_m\,\partial_{n_1}\,q_{\alpha\beta}\,A^{\alpha\beta}_{n_2\cdots
n_9}\right) -\int_{{{\mathcal M}_8}}\,d^{8}\xi\,\,T
\,\sqrt{-g_{(8)}}- \int_{{{\mathcal M}_8}}\,
q_{\alpha\beta}\,A^{\alpha\beta}_{(8)}\,, \nonumber
\end{eqnarray}
with $F_m$ being an auxiliary vector field. We derive from this
action the equations of motion for $F_m$ and
$q_{\alpha\beta}\,A^{\alpha\beta}_8$ (which only appears in the
first and the third term of the second line):
\begin{equation}\label{f1}
d\,F_{1}=-{}^*J_{8}=-d{}^*G_{9}\quad\Rightarrow\quad
F_{1}=:\,(d\chi'-{}^*G_{9})\,,
\end{equation}
and
\begin{equation}\label{solution}
{~}^*d\,(q_{\alpha\beta}\,A^{\alpha\beta}_{8})={~}^*q_{\alpha\beta}F^{\alpha\beta}_{(9)}=\left(T^2-4{{{\rm
det}~Q}}\right)\,F_{1}\,.
\end{equation}
These two equations implicitly define the axion $\chi'$. Now
substituting the solution (\ref{f1}) for $F_{1}$ back into the
action (\ref{ad+7+10}) we get the analogue of \eqref{ad+D7-WZ}
\begin{eqnarray}\label{ad+7}
 S_{\text{I}}&=&\int_{\mathcal
M_{10}}\,d^{10}x\,\sqrt{-g_{(10)}}\,\left(R-{{{\partial_m}T{\partial^m}T}\over
{2\left(T^2-4{\rm det}~Q\right)}}\,-{1\over 2}\left(T^2-4{\rm
det}~Q\right)\,F_m F^m\right) \cr &&-\int_{\mathcal
M_{8}}d^8\xi\,\,T\,\,\sqrt{-g_{(8)}} \,,
\end{eqnarray}
where now $F_{1}=dx^mF_m=(d\chi'-{}^*G_{9})$ and the Wess--Zumino
term has disappeared.

On the other hand instead of using Eq. \eqref{f1} we can
substitute in \eqref{ad+7+10} the solution for $F_{1}$ in terms of
the dual field strength ${~}^*q_{\alpha\beta}F_{9}^{\alpha\beta}$
\eqref{solution} and obtain the local action describing the minimal
coupling of the Q7-brane to the
$SL(2,\mathbb{R})$ invariant fields $T$ and
$q_{\alpha\beta}\,A^{\alpha\beta}_{8}$
\begin{eqnarray}\label{ad+78}
S_{\text{II}}=\int_{\mathcal
M_{10}}\,d^{10}x\,\sqrt{-g_{(10)}}\,\left[R-{{1}\over
{2\left(T^2-4{\rm
det}~Q\right)}}\,\Big({\partial_m}T{\partial^m}T\right.\hspace{80pt}
\nonumber\\
\\
\left.+ \frac{1}{9!}\, q_{\alpha\beta}F_{m_1\cdots
m_9}^{\alpha\beta}q_{\gamma\delta}F^{\gamma\delta\,\,m_1\cdots
m_9}\Big)\right] -\int_{\mathcal M_{8}}d^8\xi\,\,T\,\sqrt{-g_{(8)}}-
\int_{\mathcal M_{8}}\, q_{\alpha\beta}A^{\alpha\beta}_{8} \,.
\nonumber
\end{eqnarray}

\subsection{Unobservability of the Dirac 8-brane}

To describe the magnetic coupling of the 7-brane to the axion we
have introduced into the actions \eqref{ad+D7-WZ} and \eqref{ad+7} the Dirac 8-brane \eqref{sarj1} and
\eqref{sarj2}. As in the classical Dirac monopole problem, the Dirac
brane is not a physical object, i.e. the dynamics of the system
should not depend on the orientation of the Dirac 8-brane in the
bulk. This is reflected in the fact that the 8-brane equations of
motion are not independent. They are satisfied provided the axion
field equations hold. To see this, let us derive the axion field
equation and the equation of motion of the embedding coordinates of
the Dirac 8-brane.

The field equation of $\chi'$ is
\begin{equation}\label{chi'}
\partial_m\,\left[\sqrt{-g_{(10)}}\,\left(T^2-4{\rm
det}~Q\right)\,F^m\right]=0\,.
\end{equation}
It is accompanied by the Bianchi identity which follows from the
definitions \eqref{f1} and \eqref{sarj1}
\begin{equation}\label{Bian-chi'}
d\,F_{1}=-{}^*J_{8}
\end{equation}
with $J_{8}$ being the 7-brane current \eqref{j}. Eqs.
\eqref{chi'} and \eqref{Bian-chi'} describe the magnetic coupling of
the 7-brane to $\chi'$.

The variation of \eqref{ad+7} with respect to the Dirac 8-brane
world-volume coordinates $\hat x^m(y)$, appearing in \eqref{f1} and
\eqref{sarj1}, produces the equation
\begin{equation}\label{chi'8}
\partial_m\,\left(\sqrt{-g_{(10)}}\,\left(T^2-4{\rm
det}~Q\right)\,F^m\right)|_{\mathcal M_9}=0\,,
\end{equation}
which is nothing but the field $\chi'$ equation \eqref{chi'} pulled
back on the Dirac 8-brane world-volume. Therefore, the Dirac brane
is not physical. Its position can be anywhere in space-time and it
is invisible provided the Dirac veto holds, which does not allow the
Dirac brane to intersect the world-volumes of the objects coupled to
$\chi'$ in an electric way. If such objects (which would be
instantons) are present, their `currents' contribute to the right
hand side of Eq. \eqref{chi'}, while Eq. \eqref{chi'8}
remains sourceless. The two equations are then consistent provided
the world-volumes of the Dirac brane and the electrically charged
objects never intersect\footnote{Additional complications and
subtleties regarding the Dirac branes and corresponding singular
terms in the action and equations of motion arise when the action
contains Wess--Zumino terms with `bare' electric and/or magnetic
potentials. In such cases it becomes much less trivial to reconcile
the Dirac veto with the physical field equations. This happens for example
in the case of the M5-brane \cite{Bandos:1997gd}. In
\cite{Lechner:1999ga} a consistent method was developed to resolve
these problems and related problems of anomalies.}. In quantum
theory, as is well known, the unobservability of the Dirac branes is
guaranteed by the Dirac quantization condition which results in the
quantization of corresponding fluxes.

\subsection{Field equations and static 7-brane
solutions}\label{statics}

Let us now consider the complete set of equations of motion which
follow from the action \eqref{ad+7}. The variation with respect to
the 10-dimensional metric results in the Einstein equations with
the energy-momentum tensor $T_{mn}$ having the contributions from
the axion-dilaton and the 7-brane
\begin{equation}\label{einstein}
R_{mn}-{1\over 2}\,g_{mn}\,R=T_{mn}\,.
\end{equation}
The variation with respect to the field $T$ gives
\begin{equation}\label{T}
{\mathcal D}_m\,\left({1\over {T^2-4{\rm
det}~Q}}\partial^m\,T\right)+{{T{\partial_m}T{\partial^m}T}\over
{\left(T^2-4{\rm
det}~Q\right)^2}}-T\,F_m\,F^m={1\over{\sqrt{-g_{10}}}}\,\int d^8\xi
\sqrt{-g_8}\,\delta(x-\hat x(\xi))\,.
\end{equation}
The field equation of $\chi'$ and the corresponding Bianchi identity
have been given, respectively, in Eqs. \eqref{chi'} and
\eqref{Bian-chi'}.

The 7-brane equation of motion obtained by varying \eqref{ad+7}
with respect to the world-volume field $\hat x^m(\xi^\mu)$ is
\begin{eqnarray}\label{7brane}
T {\mathcal D}_\mu\left(\sqrt{-g_8}\,g^{\mu\nu}_8\partial_\mu\,\hat
x^n\right)\,g^{10}_{nm}-\sqrt{-g_8}\,\partial_\mu T {\hspace{200pt}}\nonumber\\
\\
={1\over 8!}\left(T^2-4{\rm
det}~Q\right)\,\epsilon^{\mu_1\cdots\mu_8}\,\partial_{\mu_1}x^{n_1}\cdots\partial_{\mu_8}x^{n_8}\,
\epsilon_{n_1\cdots n_8ml}\,F^l(\hat x(\xi))\,,\nonumber
\end{eqnarray}
where $\mu,\nu=0,1,\cdots,8$ are 7-brane world-volume indices and
${\mathcal D}_\mu$ is the pullback of the 10D bulk covariant
derivative containing the Christoffel symbols.

Let us now consider supersymmetric solutions of Eqs.
\eqref{einstein} to \eqref{7brane} corresponding to the 7-branes.
The simplest possible assumption is that the 7-brane is static and
does not fluctuate in the transverse directions
$x^{8,9}=\rm{cst}$. We can also assume that the axion and dilaton
fields depend only on the transverse coordinates $x^i$ $(i=8,9)$.
Such an Ansatz corresponds to the dimensional reduction of the
supergravity -- 7-brane system to a 2-dimensional system with the
7-brane being `shrunk' to a point in the 2-dimensional space. The
Dirac 8-brane then reduces to a Dirac string which ends on the
pointlike particle counterpart of the 7-brane.

The consistency condition, which must hold in order that the 7-brane
is static, is obtained by using the static gauge and setting in
\eqref{7brane} the derivatives of the transverse scalars $\hat x^8$
and $\hat x^9$ equal to zero. It has the form
\begin{equation}\label{static}
    \partial_{i}T=-(T^2-4\rm{det}\,Q)\sqrt{-g}\epsilon_{01\ldots
    7ij}(\partial^{j}\chi'-(*G_{(9)})^j)\,,
\end{equation}
where $i,j=8,9$. Outside the source using that the metric
longitudinal to the 7-brane is flat (which follows from
supersymmetry) we find
\begin{equation}\label{static2}
    \partial_{8}T=-(T^2-4\rm{det}\,Q)\partial_{9}\chi'\,,\quad\partial_{9}T=(T^2-4\rm{det}\,Q)\partial_{8}\chi'\,.
\end{equation}

These equations can be used to construct the solution for $T$ and $\chi'$
in the neighborhood of a 7-brane, i.e. near $z=0$.
Since the axion describes the magnetic charge of the 7-brane, i.e. the axion charge equals
\begin{equation}
    m=\int_0^{2\pi}\frac{d\chi'}{d\theta}d\theta\,
\end{equation}
with $z=x^8+ix^9=re^{i\theta}$, near the 7--brane we can take
\begin{equation}
    \chi'=\frac{m}{2\pi}\,\theta\,.
\end{equation}
Writing equations \eqref{static2} in the form of the Cauchy--Riemann
differential equations we find that for $\text{det}\,Q=0$ near $z=0$
\begin{equation}\label{solutiondetQ=0}
    \tau=\chi'+\frac{i}{T}=\frac{m}{2\pi i}\log z
\end{equation}
while for $\text{det}\,Q>0$ near $z=0$ we have
\begin{equation}\label{mathcalT}
    \mathcal{T}=\chi'+\frac{i}{4\sqrt{\rm{det}\,Q}}\log\frac{T+2\sqrt{\rm{det}\,Q}}{T-2\sqrt{\rm{det}\,Q}}=
\frac{m}{2\pi i}\log z\,.
\end{equation}
In both cases, solutions \eqref{solutiondetQ=0} and \eqref{mathcalT}, the axion charge $m$ is equal to the number of 7-branes which are located at the point $z=0$.

In footnote \ref{negativetension} it is mentioned that for $q<0$ the
tension of a 7-brane can be negative and that negative tension
Q7-branes are used in the construction of globally well-defined
7-brane solutions. So we would like to include them as potential
source terms to the IIB supergravity action. It can be seen from Eq.
\eqref{mathcalT} that near $z=0$ the tension $T$ of a Q7-brane will
be negative whenever $m<0$. We will however prefer to keep $m>0$ so
that for a positive/negative tension Q7-brane we have
\begin{equation}\label{mathcalT2}
    \text{sign}(q)\,\mathcal{T}=\frac{m}{2\pi i}\log z\,.
\end{equation}

The condition \eqref{static} under which the 7-brane can be
considered static coincides with the condition that $\tau$ or
$\mathcal{T}$ are holomorphic functions with logarithmic branch
cuts. As it follows from equation \eqref{static}, along the Dirac
string the holomorphicity fails, so the Dirac string plays the
role of a branch cut of $\tau$ or $\mathcal{T}$. Crossing of these
branch cuts is related to the nontrivial monodromy of the
functions $\tau$ and $\mathcal{T}$. This will be the subject of
the next Section.

\setcounter{equation}0

\section{Dirac strings and monodromy}\label{sec:Diracstrings}

\subsection{Q7-branes in the $(\tau,\bar\tau)$ basis}

Let us now consider how a Q7-brane, magnetically coupled to the
field $\chi'$, couples to the conventional
axion and dilaton $(\chi,\phi)$ or rather to $(\tau,\bar\tau)$. The
duality relations \eqref{f1}, \eqref{solution} and \eqref{duality}
(taken in the absence of the Dirac brane) prompt us that the
differential of $\chi'$ is expressed in terms of the $SU(1,1)/U(1)$
Cartan forms (\ref{defP}) and (\ref{defbarP}) as follows
\begin{equation}\label{SU(1,1)axion}
    (T^2-4\text{det}Q)\,d\chi'=-iq_{\alpha\beta}\left(V_{-}^{\alpha}V_{-}^{\beta}P-
    V_{+}^{\alpha}V_{+}^{\beta}\bar P\right)=q_{\alpha\beta}\,^*F_{9}^{\alpha\beta}\,.
\end{equation}
We substitute this expression for $d\chi'$ into the action
\eqref{ad+7} and subsequently use Eq. \eqref{tension} and Eqs.
\eqref{parametrization} to \eqref{Pintermsoftau} of Appendix
\ref{coset}. This yields the following action for the coupling of
the Q7-brane to the standard form of the IIB supergravity action
\begin{align}\label{couplingintaubasis}
    S=&\int_{\mathcal{M}_{10}}d^{10}x\sqrt{-g_{(10)}}\left[R-
    \frac{1}{2(\text{Im}\tau)^2}\vert\partial_{m}\tau+(p+q\tau^2+r\tau)\left(*G_{9}\right)_{m}\vert^2\right]\nonumber \\
    &-\int_{\mathcal{M}_{8}}d^8\xi\frac{1}{\text{Im}\tau}(p+q\vert\tau\vert^2+r\frac{\tau+\bar\tau}{2})
    \sqrt{-g_{(8)}}\,.
\end{align}
This action also applies when $\text{det}Q=0$. For the case $p=1$
and $q=r=0$ it can be seen to coincide with \eqref{ad+D7-WZ}.

Another way to write the action \eqref{couplingintaubasis} is as
follows
\begin{equation}
    S=\int_{\mathcal{M}_{10}}d^{10}x\sqrt{-g_{(10)}}\left(R-2\hat P_{m}^{*}\hat P^m\right)
    -\int_{\mathcal{M}_{10}}d^{10}x\int_{\mathcal{M}_{8}}d^8\xi\,\delta(x-\hat x(\xi))
    q_{\alpha\beta}V^{\alpha}_{-}V^{\beta}_{+}
    \sqrt{-g_{(8)}}\,,
\end{equation}
where $\hat P_{m}$ is
\begin{equation}\label{defPhat}
    \hat
    P_{m}=P_{m}-\frac{i}{2}q_{\alpha\beta}V_{+}^{\alpha}V_{+}^{\beta}\left(*G_{9}\right)_{m}\,.
\end{equation}
In terms of $\hat P$ the duality relation between the 8-form and
$\chi'$ field strengths in the presence of sources takes the form
\begin{equation}
    q_{\alpha\beta}F^{\alpha\beta}_{9}=-i*\left(q_{\alpha\beta}V_{-}^{\alpha}V_{-}^{\beta}\hat P-
    q_{\alpha\beta}V_{+}^{\alpha}V_{+}^{\beta}\hat P^{*}\right)\,.
\end{equation}
This relates the equations of motion and the Bianchi identity of the
8-form $q_{\alpha\beta}A^{\alpha\beta}_{8}$ to the Bianchi identity
and the equations of motion of the axion $\chi'$.

The Bianchi identity for $\hat P_{m}$ can be written as
\begin{equation}\label{BianchiP}
    \hat D\hat P=: d\hat P-2i\hat{\mathcal
    Q}\wedge \hat
    P=-\frac{i}{2}q_{\alpha\beta}V_{+}^{\alpha}V^{\beta}_{+}*J_{8}\,,
\end{equation}
where
\begin{equation}\label{defQhat}
    \hat
    {\mathcal Q}_{m}={\mathcal Q}_{m}+\frac{1}{2}q_{\alpha\beta}V_{+}^{\alpha}V_{-}^{\beta}\left(*G_{9}\right)_{m}\,
\end{equation}
and ${\mathcal Q}_{m}$ is a composite $U(1)$ gauge field defined in
\eqref{defQ}. Eq. \eqref{BianchiP} is the `sourced' version
of the Bianchi identity \eqref{Bianchi}, i.e. of $DP=dP-2i{\mathcal Q}\wedge P=0$.

\subsection{Monodromy}\label{monodromy}

The hatted 1-form fields $\hat P$ and $\hat {\mathcal Q}$ defined
in \eqref{defPhat} and \eqref{defQhat} can be collected into the
matrix-valued 1-form
\begin{equation}\label{hattedfields}
     V\left(\begin{array}{cc}
    -i\hat {\mathcal Q} & \hat P \\
    \overline{\hat P} & i\hat {\mathcal Q}
    \end{array}\right)V^{-1}=dVV^{-1}+SQS^{-1}*G_{9}\,,
\end{equation}
with $S$ defined in Eq. \eqref{matrixQ} and where $V$ is the matrix defined in equation \eqref{matrixV}.
Let us define
\begin{equation}
    \hat p=p+SQS^{-1}*G_{9}\,,
\end{equation}
where
\begin{equation}\label{defp}
    p=V\left(\begin{array}{cc}
                            0 & P \\
                            \bar P & 0
                            \end{array}\right)V^{-1}\,.
\end{equation}
Then $\hat p$ satisfies the Bianchi identity
\begin{equation}\label{Bianchiphat}
    d\hat p-2p\wedge p=SQS^{-1}*J_{8}\,.
\end{equation}
Outside the source the Bianchi identity \eqref{Bianchiphat} is
solved by
\begin{equation}
p = \frac{1}{2}(dC)C^{-1}\,,
\end{equation}
where $C=V{V^{\dagger}}^{\,}$\footnote{The relation between the
matrix $C$ and the matrix
%\begin{equation}
$ M=
    e^{\phi}\left(\begin{array}{cc}
    \vert\tau\vert^2 & \chi \\
    \chi & 1
    \end{array}
    \right)
$
%\end{equation}
is given by $M=S^{-1}CS$ with $S$ given in \eqref{matrixQ}.}.
Alternatively, we can write this solution as
\begin{equation}\label{paralleltransportC}
DC=0\hskip 1truecm {\rm with}\hskip 1truecm D=d-2p\,.
\end{equation}
This equation can be interpreted as saying that $C$ is parallel
transported with respect to the flat connection $p$. Let
$\gamma(\lambda)$ be some path parameterized by $\lambda$ which runs
from 0 to 1. Then we have
\begin{equation}
    C(\lambda=1)=\mathcal{P}\,\text{exp}[2\int_{\gamma}p]\,\;C(\lambda=0)\;,
\end{equation}
where $\mathcal{P}$ denotes the path ordering symbol. Since the
connection is flat the quantity
$\mathcal{P}\,\text{exp}[2\int_{\gamma}p]$ for closed $\gamma$
will only depend on the base point of the closed path. The location
of the base point can be changed by a similarity transformation,
\begin{equation}
    \mathcal{P}\text{exp}[2\oint_{\gamma}p]~\rightarrow ~H\,\mathcal{P}\text{exp}[2\oint_{\gamma}p]\,H^{-1}\quad
    \text{where}\quad H=\mathcal{P}\text{exp}[2\int_{\tilde\gamma}p]\;,
\end{equation}
with the path $\tilde\gamma$ connecting the initial to the final
base point. This means that the eigenvalues of the monodromy matrix,
$\mathcal{P}\text{exp}[2\oint_{\gamma}p]$, are preserved under
shifting the position of the base point. Therefore a physical quantity that we can associate with the Bianchi
identity $dp-2p\wedge p=0$ is the Wilson line\footnote{The terminology
is borrowed from Yang--Mills theory. Here $p$ is not a gauge field.
It is because of a mathematical similarity that we call this
quantity a Wilson line.}
\begin{equation}\label{Wilsonline}
    \text{Tr}\,\mathcal{P}\,\text{exp}[2\oint_{\gamma}p]\,.
\end{equation}
For 7-brane solutions the matrix $p$ only depends on the two
coordinates transverse to the brane. In that case the quantity
$\mathcal{P}\text{exp}[2\oint_{\gamma}p]$ will determine the
monodromy of $C$ and thus of the scalars which parameterize it.

The monodromy of the matrix $C=VV^{\dagger}$ is given by
\begin{equation}
    C(\lambda=1)=\mathcal{P}\,\text{exp}[2\oint_{\gamma}p]\,\;C(\lambda=0)\,.
\end{equation}
Let us consider a path $\gamma$ which encircles the 7-brane (point)
source in the transverse space. Further we assume that $\gamma$
encloses an area of infinitesimal size, denoted by $D$. Expanding
the path-ordered expression up to second order we find
\begin{equation}
    C(\lambda=1)=\left(1+\int_{D}\left(d\hat p-2p\wedge p\right)+
    \ldots\right)C(\lambda=0)\left(1+\int_{D}\left(d\hat p-2p\wedge p\right)^{\dagger}\right)\,,
\end{equation}
where we have used the fact that $pC=Cp^{\dagger}$ and
$\oint_{\gamma} p=\int_{D}d\hat p$. According to Eq.
\eqref{Bianchiphat} this can be written as
\begin{equation}
    C(\lambda=1)=\left(1+SQS^{-1}+\ldots\right)C(\lambda=0)\left(1+(SQS^{-1})^{\dagger}+\ldots\right)\,.
\end{equation}
Since we know the monodromy of $C=VV^{\dagger}$ when going at an
infinitesimal distance around a 7-brane and since we know the
parametrization of $V$ in terms of $\tau$, see Eq.
\eqref{parametrization}, we know the monodromy of $\tau$. It follows
that $\tau$ transforms as
\begin{equation}
    \tau~\rightarrow ~e^{Q}\tau\,.
\end{equation}

The Wilson line \eqref{Wilsonline} when evaluated around the contour $\gamma$ encircling a 7-brane at an infinitesimal distance can be evaluated and is equal to $\text{Tr}\,e^Q$. Hence, the Wilson line computes what is called the $SL(2,\mathbb{R})$ conjugacy class (see Subsection \ref{cosys} for more details about the $SL(2,\mathbb{R})$ conjugacy classes).

%In order to evaluate the Wilson line
%\begin{equation}\label{Wilsonline}
% \text{Tr}\mathcal{P}\,\text{exp}[2\oint_{\gamma}p]
%\end{equation}
%up to the first nontrivial term we use the following general
%result\footnote{Taken from Peskin and Schroeder}. Let $A(\epsilon)$
%be an arbitrary $SU(1,1)$ matrix which tends to 1 as
%$\epsilon\rightarrow 0$ then we can write $A$ as
%\begin{equation}
% A(\epsilon)=\exp\left[\epsilon\alpha_{1}^{i}T^{i}+\epsilon^2\alpha_{2}^{i}T^{i}+\cdots\right]\,,
%\end{equation}
%where the $T^{i}$ are the generators of the $SU(1,1)$ group and
%where the $\alpha^{i}_{1}$ etc. are arbitrary functions. Expanding
%$A(\epsilon)$ up to second order we see that the trace of $A$ up to
%this order in $\epsilon$ is given by
%\begin{equation}
% \text{Tr}A(\epsilon)=2+\frac{1}{2}\text{Tr}\left(\epsilon\alpha_{1}^{i}T^{i}\right)^2+\cdots\,.
%\end{equation}
%Applying this formula to equation \eqref{Wilsonline} with
%\begin{equation}
% \epsilon\alpha_{1}^{i}T^{i}=2\int_{D}\left(d\hat p-2p\wedge
% p\right)
%\end{equation}
%we obtain the following result
%\begin{equation}
% \text{Tr}\mathcal{P}\,\text{exp}[2\oint_{\gamma}p]=2+\frac{1}{2}\text{Tr}\,4(SQS^{-1})^{2}+\cdots=2-4\text{det}\,Q+
% \cdots\,.
%\end{equation}
%We thus conclude that the Wilson line, equation \eqref{Wilsonline},
%computes the monodromy class.

It was mentioned that in general the monodromy of the matrix $C$ and
thus of $\tau$ is base point dependent. When $\tau$ is an analytic
function (as in the case of 7-brane solutions) with a given
monodromy around some closed contour $\gamma$, it must have a branch
cut. In this case the base point dependence relates to the ordering
of these branch cuts.

%\subsection{A pseudo action}

%\textbf{The approach taken in \cite{Bergshoeff:2006jj} is based on
%the equivalence between Dirac strings and branch cuts. If one
%restricts to those field configurations which are 1/2 BPS, i.e.
%those for which $\tau$ is a holomorphic function then one can take
%action \eqref{multiple} without the terms proportional to
%$*G_{(9)}$. However, the resulting action is a pseudo action in
%that it can only be used for 1/2 BPS field configurations.}

%\section{Coordinate systems on $SL(2,\mathbb{R})/SO(2)$}\label{coordinatesystems}

\subsection{Multiple 7-branes}

The action describing the coupling to the IIB supergravity
axion-dilaton sector of $n$ 7-branes is given by
\begin{equation}\label{multiple}
    S=\int_{\mathcal{M}_{10}}d^{10}x\sqrt{-g_{(10)}}\left(R-2\hat P_{m}^{*}\hat P^m\right)
    -\int_{\mathcal{M}_{10}}d^{10}x\sum_{k=1}^{n}\int_{\mathcal{M}^{k}_{8}}d^8\xi_k\,\delta(x-\hat x_k(\xi_k))
    q^{k}_{\alpha\beta}V^{\alpha}_{-}V^{\beta}_{+}
    \sqrt{-g^{k}_{(8)}}\,,
\end{equation}
with
\begin{equation}\label{defPhatm}
    \hat
    P_{m}=P_{m}-\frac{i}{2}\sum_{k=1}^{n}q^{k}_{\alpha\beta}V_{+}^{\alpha}V_{+}^{\beta}\left(*G^{k}_{9}\right)_{m}\,.
\end{equation}
The world-volume $\mathcal{M}^{k}_{8}$ of each 7-brane, carrying a
charge $q^{k}_{\alpha\beta}$, is parameterized by $\xi_{k}^{\mu}$
and is located in target space at the point $\hat x_k(\xi_k)$. Its
embedding metric is $g^{k}_{\mu\nu}$ and the Dirac 8-brane
stemming from the 7-brane is described by $G^{k}_{ 9 }$.

When Dirac strings stemming from 7-branes with different charge matrices $Q$ intersect there will generically be a nontrivial monodromy for $\tau$ when going around the intersection point. The intersection point of two Dirac strings however does not describe the locus of another 7-brane, so we demand that Dirac strings can only intersect when the total monodromy measured when going around the intersection point is the identity.

\begin{figure}[h]
    \begin{center}
        \begin{pspicture}(0,0)(8,5)
            \psset{unit=.6cm}
            \psline[linewidth=.5pt](1,1)(7,4)
            \psline[linewidth=.5pt](1,4)(7,1)
            \psdot(4,2.5)
            \psarc[linewidth=.5pt](4,2.5){.7}{0}{180}
            \psarc[linewidth=.5pt,arrows=->](4,2.5){.7}{180}{360}
            \psdot(1,1)\psdot(1,4)
            \pscurve[linewidth=.5pt,arrows=->](1.3,.7)(1.5,1,2)(1.3,1.7)
            \pscurve[linewidth=.5pt,arrows=->](1.2,3.4)(1.46,3.8)(1.41,4.3)
            \uput[u](4,3.5){$b$} \uput[ur](1.3,1.7){$e^{Q_2}$}
            \uput[ur](1.41,4.3){$e^{Q_1}$}\uput[r](5,2.5){$e^{Q_2}e^{Q_1}e^{-Q_2}e^{-Q_1}$}
        \end{pspicture}
    \end{center}
    \caption{Two intersecting Dirac strings stemming from 7-branes with charge matrices $Q_1$ and $Q_2$. The point $b$ is taken as the base point for the monodromy of $\tau$ around the intersection point.}\label{twointersectingDS}
\end{figure}
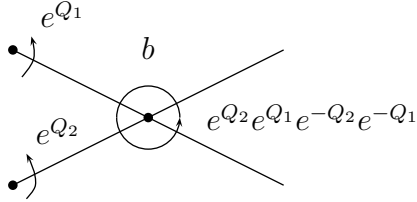

Consider figure \ref{twointersectingDS} which shows two intersecting Dirac strings stemming from different 7-branes characterized by the charge matrices $Q_1$ and $Q_2$. The point $b$ is taken as the base point to evaluate the monodromy of $\tau$ when going around the intersection point. The measured monodromy around the intersection point is
\begin{equation}\label{intersectionpoint}
    e^{Q_2}e^{Q_1}e^{-Q_2}e^{-Q_1}\,.
\end{equation}
Since there is no brane located at the intersection point by
assumption we must have that the monodromy \eqref{intersectionpoint}
is equal to identity. This is only possible when
\begin{equation}\label{intersectioncondition}
    [Q_1,Q_2]=0\,.
\end{equation}
It can be seen that the condition \eqref{intersectioncondition} is
base point independent. Hence, the Dirac strings of two 7-branes
whose charge matrices are not proportional to each other
($Q_1\neq\alpha Q_2$) are not allowed to intersect.

The fact that the Dirac strings of two 7-branes for which
$Q_1\neq\alpha Q_2$ are not allowed to intersect each other is
potentially worrisome because the Dirac strings are defined on the
2-dimensional transverse space and hence always intersect each
other, if not at some finite point then at infinity. However, it
is possible for three different Dirac strings to intersect each
other as we will discuss next.

Consider the case of three different 7-branes with charges $Q_1$,
$Q_2$ and $Q_3$ such that
\begin{equation}
    [Q_1,Q_2]\neq 0\,,\qquad [Q_1,Q_3]\neq 0\,,\qquad [Q_2,Q_3]\neq 0\,,
\end{equation}
but which satisfy
\begin{equation}
    e^{Q_1}e^{Q_2}e^{Q_3}=\mathbbm{1}\,.
\end{equation}
Then it is allowed for the collection of all three Dirac strings to
intersect each other at one point. This property of three different
Dirac strings is the basis of the construction of the globally
well-defined 7-brane solutions of \cite{Bergshoeff:2006jj} where it is shown that any 7-brane configuration can be
obtained by taking combinations of, what are referred to as, the 1A and 1B buidling blocks. Each of these building blocks consists of three 7-branes.

%Another question which one may ask is: when is it allowed to put
%different 7--branes with different charge matrices on top of each
%other? A stack of coincident 7--branes can be formed whenever the
%collection of all the Dirac strings of the individual branes are
%allowed to intersect in one point. When this is the case we can move
%all Dirac strings on top of each other forming a single Dirac string
%of the stack. We conclude that we cannot, for example, construct two
%coincident 7-branes for which $Q_1\neq\alpha Q_2$. {\bf [TO
%DISCUSS]}

\setcounter{equation}0

\section{Conjugacy class dependence of IIB supergravity}\label{sec:conjugacyIIB}

In Section \ref{c8chi'} the scalars $T$ and $\chi'$ were introduced.
In this Section we will discuss in detail the geometrical nature of
the relation between the two sets of the scalar fields
$(\tau,\bar\tau)$ and $(T,\chi')$. The IIB action in terms of $T$
and $\chi'$, eq. \eqref{ad+7} (without source terms), contains the
parameter $\text{det}\,Q$, which labels the $SL(2,\mathbb{R})$
conjugacy classes. As long as the action is written in terms of
scalars and not in terms of 8-forms the parameter $\text{det}\,Q$
can be transformed away by an appropriate field redefinition. When
the axion $\chi'$ has been dualized into the 8-form
$q_{\alpha\beta}A_8^{\alpha\beta}$ it is no longer possible to
transform away the conjugacy class label $\text{det}\,Q$, i.e. we
cannot by means of a local field redefinition go from the action
\eqref{ad+78} (without source terms) to the action
\eqref{actionII} (without source terms). Whenever the conjugacy
class label $\text{det}\,Q$ cannot be transformed away by (local)
field redefinitions we say that the resulting system is manifestly
conjugacy class dependent. Of course another obvious situation in
which one cannot transform away the parameter $\text{det}\,Q$ is
when one couples a Q7-brane to the IIB action.

\subsection{Coordinate systems on the scalar manifold}\label{cosys}

>From Eq. \eqref{ad+7} it follows that in the absence of the 7-brane
sources the kinetic term for the scalar fields $\chi'$ and $T$ has
the form
\begin{equation}\label{Tchi'kinetic}
    \mathcal{L}_{\text{scalar KT}}=
    -\sqrt{-g}\left(\frac{1}{2}\frac{1}{T^2-4\text{det}\,Q}\,\partial_m T\partial^m T+\frac{1}{2}(T^2
    -4\text{det}\,Q)\,\partial_m\chi'\partial^m\chi'\right)\,.
\end{equation}
The scalar field kinetic terms can be read as a line element of the
space $SL(2,\mathbb{R})/SO(2)$. This is a maximally symmetric space
which has three Killing vectors. The Killing vectors are
differential operators which generate the Lie algebra of
$SL(2,\mathbb{R})$. Along an integral curve generated by a Killing
vector the metric does not change. In an adapted coordinate system
one of the coordinates, here $\chi'$, of the line element runs along
such an integral curve. The presence of the Killing symmetry
associated with the shift of $\chi'$ is expressed by the fact that
the metric components do not depend on
$\chi'$. We thus see that the shift symmetry of $\chi'$ in adapted
coordinates corresponds to an $SL(2,\mathbb{R})$ transformation
$\tau \rightarrow e^Q\, \tau$ of the axion-dilaton field.

In Eq. \eqref{mathcalT} we introduced a new complex field
$\mathcal{T}$. In terms of this complex field the scalar field kinetic part
of the IIB supergravity action \eqref{Tchi'kinetic} takes the form
\begin{equation}\label{mathcalTK}
    \mathcal{L}_{\mathcal T}=-\sqrt{-g}\,\frac{1}{2}\,\frac{\text{det}\,Q\,\partial_m
    \mathcal{T}\partial^{m}\bar{\mathcal{T}}}{4\sinh^2\,(2\sqrt{\rm{det}\,Q}\,{\rm{Im}\mathcal{T}})}
\end{equation}
while in terms of $\tau$ it is given by
\begin{equation}\label{tau}
    \mathcal{L}_{\tau}=-\sqrt{-g}\,\frac{1}{2}\,\frac{\partial_{m}\tau\partial^{m}\bar\tau}{(\rm{Im}\,\tau)^2}\,.
\end{equation}
Comparing \eqref{mathcalTK} and \eqref{tau} one finds the following
relation between
$\tau$ and $\mathcal{T}$ \footnote{The relation between $\tau$ and $\mathcal{T}$ could also have been obtained from the local solutions for $\tau$ and $\mathcal{T}$ near a Q7-brane. From Eq. \eqref{mathcalT2} we know that $z=e^{2\pi i\,\text{sign}(q)\,\mathcal{T}/m}$ while from \cite{Bergshoeff:2006jj} we know that near a Q7-brane we have $z=\left(\frac{\tau-\tau_0}{\tau-\bar\tau_0}\right)^{\tfrac{\pi}{\sqrt{\text{det}\,Q}}}$ so that for $m=1$ (one Q7-brane) we obtain the relation \eqref{tautoT}.}
\begin{equation}\label{tautoT}
    w=:\frac{\tau-\tau_0}{\tau-\bar\tau_0}=e^{2i\,\text{sign}(q)\,\sqrt{\rm{det}\,Q}\,\mathcal{T}}\,,
\end{equation}
where $\tau_0$ is given by
\begin{equation}\label{fixed}
\tau_0=-\frac{r}{2q}+\frac{i}{\vert q\vert}\sqrt{\rm{det}\,Q}\,.
\end{equation}
The point $\tau_0$ is a fixed point under the
$e^Q$ transformation, i.e.
$e^Q\tau_0=\tau_0$ for $\text{det}\,Q\ge 0$~\footnote{The action of a matrix $\Lambda=\left(\begin{array}{cc}
a & b \\
c & d
\end{array}\right)$ on $\tau$, written as $\Lambda\tau$, is defined as $\Lambda\tau={{a\tau+b}\over{c\tau+d}}$.}.
Eq. \eqref{tautoT} also defines the complex scalar $w$. From Eqs. \eqref{tautoT} and \eqref{mathcalT} we find that $T$
and
$\chi'$ are given by the following expressions in terms of $\tau$
and $\tau_0$
\begin{align}
    & T=\frac{1}{\text{Im}\tau}\left(q\vert\tau\vert^2+r\text{Re}\tau+p\right)\,,\label{T1}\\
    & e^{4i\,\text{sign}(q)\,\sqrt{\text{det}\,Q}\chi'}=\frac{\vert\tau\vert^2-2\tau_0\text{Re}\tau+\tau_0^2}
    {\vert\tau\vert^2-2\bar\tau_0\text{Re}\tau+\bar\tau_0^2}\,.\label{chi'1}
\end{align}
The sign of the tension $T$ equals the sign of the parameter $q$ as follows from the following way of writing $T$ in terms of $(\tau,\bar\tau)$
\begin{equation}
    T=\frac{q}{2\,\text{Im}\tau}\left[\vert\tau-\tau_0\vert^2+\vert\tau-\bar\tau_0\vert^2\right]\,.
\end{equation}

The transformation from $\tau$ to $\mathcal{T}$ is a conformal mapping. Consider the sequence
\begin{equation}
    \tau: ({\rm upper\,\, half-plane})\longrightarrow w: (\rm{unit\,\, disk})\longrightarrow
    \mathcal{T}: (\rm{vertical\,\, strip})\,.
\end{equation}
We first map $\tau$, which takes value in the upper half plane
$(\rm{Im}\tau>0)$, to $w$ which parameterizes the unit disk $(\vert w\vert<1)$. Then we map this to
$e^{2i\,\text{sign}(q)\,\sqrt{\rm{det}\,Q}\,\mathcal{T}}$ a vertical strip in a new upper half plane $(\text{sign}(q)\,\rm{Im}\mathcal{T}>0)$.
The region $\rm{Im}\mathcal{T}>0$ corresponds to
$2\sqrt{\rm{det}\,Q}<\text{sign}(q)\,T<\infty$. The real line $\rm{Im}\tau=0$ gets
mapped to the unit circle $\vert w\vert=1$ and subsequently to the
real line $\rm{Im}\mathcal{T}=0$ which however in the process has
become periodically identified, $\chi'\sim\chi'+{\pi\over \sqrt{\rm{det}\,Q}}$. Hence, the scalar field redefinition can be read as a conformal mapping from the upper
half--plane $\text{Im}\tau>0$ to the vertical strip
$\text{sign}(q)\,\text{Im}\mathcal{T}>0$ with $\chi'\sim\chi'+{\pi\over \sqrt{\rm{det}\,Q}}$.

In the $(\mathcal{T},\bar{\mathcal{T}})$ coordinates the $SL(2,\mathbb{R})$
symmetry is no longer manifest. The global symmetries are
\begin{equation}
    e^{2i\,\text{sign}(q)\,\sqrt{\rm{det}\,Q}\,\mathcal{T}}~\rightarrow~\frac{\alpha e^{2i\,\text{sign}(q)\,\sqrt{\rm{det}\,Q}\,\mathcal{T}}
    +\beta}{\bar\beta e^{2i\,\text{sign}(q)\,\sqrt{\rm{det}\,Q}\,\mathcal{T}}+\bar\alpha}
    \qquad\rm{with}\qquad
    \vert\alpha\vert^2-\vert\beta\vert^2=1\,.
\end{equation}
It is not possible to realize the global symmetry directly on
$\mathcal{T}$. The only manifest global symmetry
left is the axion $\chi'$ shift symmetry. We have for
$\text{det}\,Q>0$ that $w$ transforms as a $U(1)$ `matter' field
\begin{equation}\label{eqtau}
    w~\rightarrow ~e^{2i\,\text{sign}\,(q)\sqrt{\text{det}\,Q}}\,w \qquad\text{when}\qquad\tau ~\rightarrow ~e^Q\tau\,.
\end{equation}
It then follows from Eq. \eqref{tautoT} that an arbitrary $\text{det}\,Q>0$ transformation $\tau \rightarrow e^Q\tau$ can be written as $\chi'\rightarrow\chi'+\text{sign}\,(q)$.

In Section \ref{wvaction} we will construct the Q7-brane Wess--Zumino and for this purpose it is convenient to introduce a complex linear combination of the RR and
NSNS 2-form fields $C_2$ and $B_2$ which transforms as \eqref{eqtau} under the action of $e^Q$. We
define the following complex 2-form $\mathcal{A}_{2}$
\begin{equation}\label{defA2}
    \mathcal{A}_{2}=:\frac{-i}{(\text{Im}\tau_0)^{1/2}}\left(-C_{2}+\tau_0 B_{2}\right)\,.
\end{equation}
Using that the 2-forms transform under $e^Q$ as (see \eqref{sltrafos})
\begin{equation}
    \left(\begin{array}{c}
    C_{2} \\
    B_{2}
    \end{array}\right)~\rightarrow ~e^Q\,\left(\begin{array}{c}
    C_{2} \\
    B_{2}
    \end{array}\right)
\end{equation}
we see that the field $\mathcal{A}_{2}$ transforms under $e^Q$ with
$\text{det}\,Q>0$ as
\begin{equation}
    \mathcal{A}_{2}~\rightarrow ~e^{i\,\text{sign}\,(q)\sqrt{\text{det}\,Q}}\,\mathcal{A}_{2}\,.
\end{equation}

With the use of Eqs. \eqref{tautoT} to \eqref{chi'1} the
duality relation \eqref{SU(1,1)axion} between the 8-form and the
axionic scalar $\chi'$ takes the form
\begin{equation}\label{axion8formduality}
 \left(T^2-4\text{det}\,Q\right)d\chi' = \frac{\text{det}\,Q}{(\text{Im}\tau_0)^2}\vert(\tau-\tau_0)(\tau-\bar\tau_0)\vert^2 d\chi'
 =*q_{\alpha\beta}F^{\alpha\beta}_9\,,
\end{equation}
where
\begin{equation}\label{defchi'}
    d\chi'=\frac{\text{Im}\tau_0}{2\sqrt{\text{det}\,Q}}\left[\frac{d\tau}{(\tau-\tau_0)(\tau-\bar\tau_0)}
    +\text{c.c.}\right]
\end{equation}
can be obtained by differentiating Eq. \eqref{tautoT}. In Appendix
\ref{sec:IIB}, Eq. \eqref{dualF9action}, we present the entire bosonic part of
IIB supergravity action in which both
$q_{\alpha\beta}F^{\alpha\beta}_9$ and $\chi'$ appear. Since in the
action \eqref{dualF9action} the parameter $\text{det}\,Q$ appears
explicitly (it also appears in $G_7$) and because it cannot be transformed away by a local field redefinition as discussed in the introduction to this Section this way of writing the IIB
supergravity action is referred to as a conjugacy class dependent
formulation.

\subsection*{The $\text{det}\,Q\rightarrow 0$ limit}

One can take at any stage the limit
$\text{det}\,Q\rightarrow 0$. Since at various places, e.g. in the
definition of $\tau_0$, Eq. \eqref{fixed}, we divide by $q$ one must
assume that $q\neq 0$. This means that when one takes the limit
$\text{det}\,Q=pq-{r^2\over 4}\rightarrow 0$ it must be assumed that
$q\neq 0$. Hence after taking this limit one ends up with a $(p,q)$
7-brane for which $q\neq 0$. In order to get to the D7-brane one
must perform an $SL(2,\mathbb{R})$ transformation which takes one
from a $(p,q)$ 7-brane to a $(1,0)$ 7-brane. The
$\text{det}\,Q\rightarrow 0$ limit of $(T,\chi')$ leads to the
expressions
\begin{align}
    & T=\frac{1}{\text{Im}\tau}\left(q\vert\tau\vert^2+r\text{Re}\tau+p\right)\,,\\
    & \chi'=-{{{r\over{2q}}+{\rm
Re}\,\tau}\over{q\,\vert\tau\vert^2+{r}\,\text{Re}\,\tau+{p}}}\,,\qquad
pq={r^2\over 4}\,.\label{qnot}
\end{align}
These two equations can be combined into the complex equation
\begin{equation}\label{detQ=0limit}
    {\cal T}=\chi'+iT^{-1}=\frac{-1}{q\tau+\tfrac{r}{2}}\,.
\end{equation}
Hence, for $\text{det}\,Q=pq-{r^2\over 4}=0$ the transformation $(\tau,\bar\tau)\rightarrow(\chi',T)$ is a
field redefinition which keeps the $SL(2,\mathbb{R})$ invariance
of the IIB supergravity action manifest, i.e. both
$\tau=\chi+ie^{-\phi}$ and ${\cal T}=\chi'+iT^{-1}$ appear in
exactly the same way in the IIB supergravity action (compare Eqs.
(\ref{mathcalTK}) and (\ref{tau}) for $\text{det}\,Q\rightarrow
0$). The reason is of course that the field redefinition in the
$\text{det}\,Q\rightarrow 0$ limit takes the form of an
$SL(2,\mathbb{R})$ transformation, Eq. \eqref{detQ=0limit}.

\subsection{What about $\text{det}\,Q<0$?}

The $SL(2,\mathbb{R})$ duality group has three subgroups:
$\mathbb{R}$, $SO(1,1)$ and $SO(2)$. The transformations $e^Q$ with
$\text{det}\,Q=0$, $\text{det}\,Q<0$ and $\text{det}\,Q>0$ belong to
these respective subgroups. In this subsection we will argue that
there are no 7-branes which correspond to the $SO(1,1)$
subgroup\footnote{The nonexistence of a 7-brane with
$\text{det}\,Q<0$ has consequences for the vacuum structure of $SO(1,1)$ gauged
supergravities in 9 dimensions.
Certain gauged 9-dimensional supergravities can be obtained by
performing a Schwerk--Schwarz reduction in which one gauges a
subgroup of $SL(2,\mathbb{R})$
\cite{Meessen:1998qm,Bergshoeff:2002mb}. This corresponds to
performing a reduction with 7-branes in the background. The fact
that there is no well-defined 7-brane for $\text{det}\,Q<0$ means that the $SO(1,1)$ 9-dimensional gauged supergravity has no well-defined domain-wall vacuum.}.

The analogue of the field redefinition \eqref{tautoT} for the case $\text{det}\,Q<0$ is
\begin{equation}\label{conformaldetQ<0}
    \frac{\tau-\tau_0^+}{\tau-\tau_0^-}=e^{2\sqrt{-\text{det}\,Q}\,\mathcal{T}}\qquad\text{where}\qquad
    \mathcal{T}=\chi'+\frac{i}{2\sqrt{-\text{det}\,Q}}\,\rm{arccot}\frac{T}{2\sqrt{-\text{det}\,Q}}\,,
\end{equation}
where
\begin{equation}
    \tau_0^{\pm}=-\frac{r}{2q}\pm\frac{1}{q}\sqrt{-\text{det}\,Q}
\end{equation}
which is such that $e^Q\tau_0^{\pm}=\tau_0^{\pm}$. The local form for $\tau$ whose monodromy around a point $z=0$ is of the form $\tau\rightarrow e^Q\tau$ with $\text{det}\,Q<0$ is given by
\begin{equation}\label{solutiondetQ<0}
    \left(\frac{\tau-\tau_0^+}{\tau-\tau_0^-}\right)^{\frac{i \pi}{\sqrt{-\text{det}\,Q}}}=z\,.
\end{equation}
The problem with this possibility is that the local solution
\eqref{solutiondetQ<0} as well as the conformal
mapping \eqref{conformaldetQ<0} are ill defined at the points
$\tau_0^{\pm}$, i.e. the limit $\tau\rightarrow\tau_0^{\pm}$ does
not exist. This means that the fixed points $\tau_0^{\pm}$ are not
part of the IIB moduli space. Indeed when we consider the moduli
space
\begin{equation}
    \frac{PSL(2,\mathbb{R})}{SO(2)\times PSL(2,\mathbb{Z})}
\end{equation}
none of the orbifold points corresponds to $\tau_0^{\pm}$.

\setcounter{equation}0
\section{Towards the construction of the Q7-brane world-volume
action}\label{wvaction}

\subsection{The Wess--Zumino term}
In previous Sections we have discussed the coupling of the Q7-brane
to the 8-form potential $q_{\alpha\beta}\,A^{\alpha\beta}_8$ and to
its magnetically dual axion field $\chi'$. In this Section we shall
construct the Wess--Zumino term which describes the coupling of the
Q7-brane to all the gauge fields of IIB supergravity. As was argued
in \cite{Bergshoeff:2006gs}, in contrast to the D7-brane and its
$SL(2,\mathbb{R})$ partners (see Eq. \eqref{DBI}), the
invariance of the Q7-brane Wess--Zumino term under the gauge
transformations (\ref{8formgauge}) requires two Born--Infeld fields
$A^\alpha_1$ $(\alpha=1,2)$ on the Q7-brane world-volume. Their field strengths are
extended with the pullbacks of the doublet of the 2-forms
$A_2^\alpha$ such that the generalized field strength
\begin{equation}\label{bifs}
\mathcal{F}^{\alpha}_2=dA_1^{\alpha}+A_2^{\alpha}=F_2^{\alpha}+A_2^{\alpha}
\end{equation}
is invariant under the gauge transformations
$$
\delta A_2^{\alpha}=d\phi_1\,,\qquad \delta A_1=-\phi_1+d\phi_0\,,
$$
where $\phi_1$ and $\phi_0$ are world-volume 1-form and 0-form gauge transformation parameters, respectively.

We demand that the Q7-brane Wess--Zumino term satisfies the following three conditions:
\begin{enumerate}
\item{it is invariant (up to a total derivative), under the gauge
transformations (\ref{8formgauge}),}
\item{it is monodromy neutral
(i.e. invariant under the $SL(2,\mathbb{R})$ transformation $e^Q$ in which $Q$ contains the charges of the 7-brane) and}
\item{it reduces in the
$\text{det}Q \rightarrow 0$ limit to the
$(p,q)$ 7-brane Wess--Zumino term of \cite{Bergshoeff:2006gs} with a single BI
field.}
\end{enumerate}
The condition that the 7-brane world-volume action must be
monodromy neutral follows from the fact that the 7-branes are always
located at fixed points of the monodromy $\tau\rightarrow e^Q\tau$.
In terms of the scalars $T$ and $\chi'$ this means that the action
must be invariant under the shift symmetry
$\chi'\rightarrow\chi'+1$ ($T$ is monodromy neutral). The WZ term
satisfying the above requirements has the following form (where for
simplicity we skip the wedge product symbol)
\begin{align}\label{Q7WZold}
&L_{\text{WZ}}=-q_{\alpha\beta}\left[A^{\alpha\beta}_8+\frac{1}{16}A_6^{(\alpha}\,
A_2^{\beta)}
   +\frac{1}{12}A_4\, A_2^{\alpha}\, A_2^{\beta}
   -\left(\frac{1}{4}A_6^{(\alpha}-\frac{1}{3}A_4\, A_2^{(\alpha}\right)\,{\mathcal F}_2^{\beta)}+
   \frac{1}{2}A_4\, {\mathcal F}_2^{\alpha}\, {\mathcal F}_2^{\beta}\right.\nonumber\\
   &\left.\hspace{50pt}+\left(\frac{1}{12}A_2^{\alpha}\, A_2^{\beta}
   -\frac{1}{4}A_2^{(\alpha}\, {\mathcal F}_2^{\beta)}\,
   +\frac{1}{4}{\mathcal F}_2^{\alpha}\,{\mathcal F}_2^{\beta}\right)\,
   \frac{i}{16}\epsilon_{\gamma\delta}A_2^{\gamma}\,
   {\mathcal F}_2^{\delta}\right]
   +a(T)\,q_{\alpha\beta}\,{\mathcal F}_2^{\alpha}\,{\mathcal F}_2^{\beta}\,{\mathcal F}_2^{\gamma}\,
   {\mathcal F}_2^{\delta}\,q_{\gamma\delta}\\
   &\hspace{30pt}+\frac{1}{6\cdot 8^3}\,(\text{det}\,Q)^{1/2}\,\left[b(T)\,e^{-4i\text{sign}(q)\,\sqrt{\text{det}\,Q}\,\chi'}
\frac{i}{(\text{Im}\tau_0)^{2}}\left(-{i}(\mathcal{F}_2^1-\mathcal{F}_2^2)
+{\tau_0}(\mathcal{F}_2^1+\mathcal{F}_2^2)\right)^4+\text{c.c.}\right]\,,\nonumber
\end{align}
where $a(T)$ and $b(T)$ are undetermined real and complex-valued
functions of $T$, respectively. The first term (in the square
brackets) is completely fixed by the requirement of gauge
invariance. In order for the WZ term
\eqref{Q7WZold} to reduce to the $(p,q)$ 7-brane Wess--Zumino term
of
\cite{Bergshoeff:2006gs} we must have that $a(T)\rightarrow$ constant
and $b(T)\rightarrow 1$ when $\text{det}\,Q\rightarrow 0$. It is
expected that the form of the functions $a(T)$ and $b(T)$ will be
fixed by world-volume supersymmetry. The last term in
\eqref{Q7WZold} describes the coupling of the Q7-brane Born--Infeld
fields to the axion $\chi'$. This term is invariant by itself under
the shift symmetry $\chi'\rightarrow
\chi'+1$ since the generalized BI field strengths
are combined into the `eigenform' of the operator $e^Q$ in a way
similar to the RR and NSNS 2-forms of Eq. \eqref{defA2}, namely
\begin{eqnarray}\label{eigenf}
e^Q:\,\frac{i}{(\text{Im}\tau_0)^{1/2}}\,\left({i}(\mathcal{F}_2^1
-\mathcal{F}_2^2)-{\tau_0}(\mathcal{F}_2^1+\mathcal{F}_2^2)\right)\rightarrow
&
\nonumber\\
&\hspace{-100pt}\rightarrow
e^{i\,\text{sign}(q)\,\sqrt{\text{det}\,Q}}\,\frac{i}{(\text{Im}\tau_0)^{1/2}}\,\left({i}
(\mathcal{F}_2^1-\mathcal{F}_2^2)
-{\tau_0}(\mathcal{F}_2^1+\mathcal{F}_2^2)\right).
\end{eqnarray}
Though, in the form presented in Eq. (\ref{Q7WZold}) the term
containing the axion $\chi'$ is not manifestly $SU(1,1)$ covariant,
it can be rewritten in an $SU(1,1)$ covariant manner, i.e. with the
scalars appearing via $V_{\pm}^{1,2}$. However, in the $SU(1,1)$
covariant form the role played by $\chi'$ is no longer manifest and
the resulting expression is more complicated. So we do not present
it here.

\subsection{The Dirac--Born--Infeld part of the Q7-brane action}

To obtain the form of the DBI part of the Q7-brane action we perform
a supersymmetry variation of the IIB background fields which appear
in the WZ term, with the parameters corresponding to the
supersymmetries which are not broken by the 7-brane and hence
leave the action invariant (for a detailed generic discussion of
this point see \cite{Bandos:2001jx}). The terms obtained by performing a supersymmetry variation of the WZ term must combine with terms which result from a supersymmetry variation of the DBI part
of the action to yield the 7-brane supersymmetry projector. It can be shown that up to terms which are cubic or
quartic in the BI field strength $\mathcal{F}_2^{\alpha}$ the supersymmetry variation of the WZ term
\eqref{Q7WZold} is given by
\begin{align}
&\delta_{\epsilon}\,S_{WZ}=-\int_{{\mathcal
M}_8}d^8\xi\left(\sqrt{-g_{(8)}}\,q_{\alpha\beta}\left[
V_-^{\alpha}V_{-}^{\beta}\bar\epsilon_C
i\gamma_{\underline{8}}\gamma_{\underline{9}}\lambda+
V_+^{\alpha}V_{-}^{\beta}\bar\epsilon\gamma_{\underline{8}}\gamma_{\underline{9}}\gamma^{\mu}\Psi_{\mu}
+\frac{1}{8}\left(V_-^{\alpha}\bar\epsilon i\gamma_{\underline{8}}\gamma_{\underline{9}}\gamma_{\mu\nu}\lambda\nonumber\right.\right.\right.\\
&\left.\left.\left.+4iV_-^{\alpha}\bar\epsilon_{C}i\gamma_{\underline{8}}\gamma_{\underline{9}}
\gamma_{[\mu}\Psi_{\nu]}\right)\mathcal{F}^{\beta\;\mu\nu}
+\frac{1}{4}V_-^{\alpha}\bar\epsilon_C\gamma_{\underline{8}}\gamma_{\underline{9}}\gamma^{\mu\nu}\gamma^{\rho}
\Psi_{\rho}\mathcal{F}_{\mu\nu}^{\beta}+\frac{1}{32}\bar\epsilon\gamma_{\underline{8}}
\gamma_{\underline{9}}\gamma^{\rho\mu\nu\sigma\tau}\Psi_{\rho}
\mathcal{F}_{\mu\nu}^{\alpha}\mathcal{F}^{\beta}_{\sigma\tau}\right]+\text{c.c.}\right)\,,\label{susyvarWZ}
\end{align}
where $\lambda$ and $\Psi_{\mu}$ are (the pullbacks of) the dilatino
and gravitino field, respectively. The supersymmetry transformations
of the IIB fields can be found in \cite{Bergshoeff:2005ac}. Greek
indices refer to world-volume indices and Latin indices to target
space indices. The underlined labels
$\underline{8}$ and $\underline{9}$ denote flat tangent space
indices. To perform the supersymmetry variation, the following gamma
matrix duality was used
\begin{equation}
-\epsilon^{\mu_1\ldots\mu_8}\gamma_{\mu_1\ldots\mu_k}=(-1)^{k(k-1)/2}k!\,\gamma^{\mu_{k+1}\ldots\mu_8}
\gamma_{\underline{0}}\gamma_{\underline{1}}\ldots\gamma_{\underline{7}}\sqrt{-g_{(8)}}\,,
\end{equation}
where $\epsilon^{\mu_1\ldots\mu_8}$ is the 8D Levi--Civit\`{a}
symbol. Further, in static gauge we have
\begin{align}
   & \gamma_{\underline{0}}\gamma_{\underline{1}}\ldots\gamma_{\underline{7}}\lambda=\gamma_{\underline{8}}
   \gamma_{\underline{9}}\lambda\,,\\
   & \gamma_{\underline{0}}\gamma_{\underline{1}}\ldots\gamma_{\underline{7}}\Psi_{\mu}=-\gamma_{\underline{8}}
   \gamma_{\underline{9}}\Psi_{\mu}\,,
\end{align}
which follows from the chirality properties $\gamma_{11}\lambda=\lambda$ and $\gamma_{11}\Psi_{i}=-\Psi_i$.
Other useful identities are the Clebsch--Gordon decompositions
\begin{align}
& \gamma_{\rho\mu\nu}=\gamma_{\mu\nu}\gamma_{\rho}+2g_{\rho[\mu}\gamma_{\nu]}\,,\\
& \gamma_{\rho\mu\nu\sigma\tau}=\gamma_{\mu\nu\sigma\tau}\gamma_{\rho}+4g_{\rho[\mu}\gamma_{\nu\sigma\tau]}\,,\label{CGdecompo}\\
& \gamma_{\mu\nu\sigma\tau}=\gamma_{\mu\nu}\gamma_{\sigma\tau}+2g_{\mu[\sigma}g_{\tau]\nu}-
2g_{\mu[\sigma}\gamma_{\tau]\nu}+2g_{\nu[\tau}\gamma_{\sigma]\mu}\,.
\end{align}

If we take for the first few terms of the DBI part of the Q7-brane
action the following one
\begin{equation}\label{DBIQ7}
  S_{DBI}=-\int_{{\mathcal M}_8}d^8\xi\,\sqrt{-g_{(8)}}\left(T+\frac{1}{4}q_{\alpha\beta}\mathcal{F}^{\alpha}_{\mu\nu}
   \mathcal{F}^{\beta\;\mu\nu}+\ldots\right)
\end{equation}
then for each term appearing in $\delta_{\epsilon}\,L_{DBI}$ there
exists a corresponding term in $\delta_{\epsilon}\,L_{WZ}$ such that
they make up the projector
\begin{equation}
    P=\frac{1}{2}(1+i\gamma_{\underline{8}}\gamma_{\underline{9}})\,.
\end{equation}
Hence, the supersymmetries which are not broken by the 7-brane are
those for which $P\,\epsilon=0$. The last two terms in the square
brackets of \eqref{susyvarWZ} are not canceled by terms in
\eqref{DBIQ7}. These terms require the modification of the projector $P$
by terms which include ${\mathcal F}_2$, analogous to those which
appear in the kappa-symmetry projector of the D7-brane.  The above
calculation generalizes the one of
\cite{Bergshoeff:2006ic} up to terms which are second order in the
BI field strength. We conclude that both Born--Infeld vectors carry
propagating degrees of freedom. In the next Section we shall briefly
discuss the possibility of reducing the number of the Born--Infeld
degrees of freedom by imposing a duality relation between their
field strengths.

\setcounter{equation}0
\section{Discussion}\label{sec:discussion}

In this paper we have considered the coupling of the Q7-brane to the
bosonic sector of IIB supergravity and the structure of its
world-volume action. In the static brane limit, in particular when
there are no BI fields on the 7-brane, the coupling is best
described using a new basis of the scalar manifold in which the
fields are $T$ and $\chi'$. The field $T$ is associated with the
tension of the Q7-brane whereas $\chi'$ is the axion dual to
$q_{\alpha\beta}A^{\alpha\beta}_8$, the 8-form to which the Q7-brane
couples electrically. Extending the construction by coupling the
Q7-brane to 0-, 2-, 4- and 6-form fields described by the gauge
invariant Wess--Zumino term requires the use of two Born--Infeld gauge
fields. At present the microscopic origin of these two BI fields and
of the Q7-brane itself is unclear. Below we only present some
speculations regarding these issues.

Since the Q7-branes preserve half of IIB D=10 supersymmetry their
$d=8$ world-volume theory is expected to possess 16 supersymmetries
and to have an equal number of bosonic and fermionic physical
degrees of freedom. In the case of the $(p,q)$ 7-branes there is a
single Born--Infeld field which in $d=8$ has 6 degrees of freedom,
two transverse scalars and 8 on-shell Goldstone fermion modes so the
number of the bosonic and fermionic degrees of freedom match. We do
not know the full bosonic Q7-brane action but the part that we do
know has in 8-dimensions a number of 14=6+6+2 bosonic degrees of
freedom (two vectors and 2 (embedding) scalars) while there are
only 8 fermionic physical degrees of freedom associated with the
Goldstone fermions of 1/2 bulk supersymmetry spontaneously broken by
the Q7-brane.

It may happen that, as in the case of the duality-symmetric
formulation of the D3-brane
\cite{Berman:1997iz,Nurmagambetov:1998gp}, the two Born--Infeld
fields are actually not independent but related to each other by a
duality condition. If a duality relation between the two
Born--Infeld fields which reduces by half their degrees of freedom
does take place, the number of bosonic and fermionic degrees of
freedom on the Q7-brane will match. A simple
proposal for such a condition which reduces by half the number of BI
physical modes, and which should probably be corrected by higher
non-linear terms (like in the case of the M5-brane
\cite{Howe:1996yn,Perry:1996mk} and the D3-brane
\cite{Berman:1997iz,Nurmagambetov:1998gp}), looks as follows
\begin{equation}\label{selfduality}
{\mathcal F}_2^-\,{\mathcal F}_2^-=i\,*({\mathcal F}_2^-\,{\mathcal
F}_2^-)\,
\end{equation}
(and the complex conjugate for ${\mathcal F}_2^+$) where the Hodge
operation is taken in the 8-dimensional worldvolume of the Q7-brane
and ${\mathcal F}_2^\pm={\mathcal
F}_2^\alpha\,V^{\beta}_\mp\,\epsilon_{\alpha\beta}$.

Another possibility is a duality relation which can be given in the
schematic form\footnote{Duality relations similar to
(\ref{selfduality}) and (\ref{1+3}) were considered e.g. in
\cite{Tchrakian:1978sf,Bais:1985ns} for studying higher dimensional counterparts of
instantonic solutions in Yang--Mills theory and their reduction to
D=4.}
\begin{equation}\label{1+3}
{\mathcal F}_2=*({\mathcal F}_2\,{\mathcal F}_2\,{\mathcal F}_2)\,.
\end{equation}
One can construct several relations of this kind by combining the
Born--Infeld field strengths with the axion-dilaton matrix
$V^{\alpha}_\mp$ and the Q7--brane charge $q_{\alpha\beta}$.

If there is no duality relation between the two BI fields then in
order to have a balance in the number of physical degrees of freedom
we need two extra scalars and 8 extra fermions. A possible
explanation for the origin of these missing degrees of freedom might
be the assumption that the Q7-brane is a bound state of two
coincident $(p,q)$ 7-branes, one of which is $SL(2,\mathbb{R})$
rotated with respect to the other, so that both e.g. the fundamental
string and the D1-brane end on the Q7-brane. If this is indeed the
case then the construction of the complete world-volume action for
the Q7-brane can be, to some extent, analogous to the construction
of the action for N coincident D-branes with non-Abelian
Born--Infeld fields
\cite{Tseytlin:1999dj,Taylor:2000pr,Myers:1999ps}. Note that the
construction of the target space covariant and supersymmetric N
D-brane action encountered serious problems which have not been
completely solved (see e.g.
\cite{Bergshoeff:2001dc,Sorokin:2001av,Drummond:2002kg,Hassan:2003uq,Howe:2005jz}
for different approaches to tackle these problems). Since in the
case under consideration we deal with only two Abelian BI fields,
one may hope that these problems can be easier overcome. There is no
non-Abelian enhancement of the gauge symmetry since there are no open
strings which connect two $(p,q)$ 7-branes whose $(p,q)$ charges
differ.

Finally one could say that if gauge invariance requires two BI
vectors then we should expect the excitations of a Q7-brane to be
always in terms of a $(p,q)$ string and a $(p',q')$ string and not
in terms of only one of them, which in turn suggests that the two
$(p,q)$ and $(p',q')$ strings are in some relation to each other and
may themselves also form a bound state. One could refer to such a
bound state as a Q1-brane. It remains to be seen if Q1-branes
require a duality relation between the two BI vectors on the
Q7-brane world-volume or not. Whether such Q1-branes exist and
whether there might be other Qp-branes which naturally couple to the
new axion-dilaton $(T,\chi')$ rather than to the conventional one is
under study.
\\
\\
{\bf Acknowledgments}. The authors are grateful to P. Argyres, I.
Bandos, M. Cederwall, J. de Azc\'arraga, T. Ort\'{\i}n, P. Pasti and
M. Tonin for useful and encouraging discussions. This work was
partially supported by the EU MRTN-CT-2004-005104 grant in which
E.B. is associated to Utrecht University and D.S. to Padova
University. J.H. was supported by a Breedte Strategie grant of the
University of Groningen. D.S. also acknowledges support from the
Ministerio de Educaci\'on y Ciencia and EU FEDER funds
(FIS2005-02761), the Generalitat Valenciana, during his sabbatical
stay in Valencia. J.H. wishes to thank the
University of Padova for its hospitality.

\appendix

\setcounter{equation}0
\section{Conventions}

We use the mostly plus signature $-+\cdots +$. The Levi--Civit\`{a}
symbol is denoted by $\epsilon_{m_1\ldots m_{10}}$ where
$\epsilon_{01\cdots 9}=-\epsilon^{01\cdots 9}=1$. We denote
space-time indices by $m,n=0,1,\ldots,9$ and 7-brane world-volume
indices by $\mu,\nu=0,1,\ldots,7$. Underlined indices refer to flat tangent space indices. Target space-time fields and world-volume fields are denoted by the same symbols.

%The embedding coordinates are
%denoted by $\hat x^{m}$ and the world-volume is parameterized by
%$\xi^{\mu}$.

\setcounter{equation}0

\section{Properties of the $SU(1,1)/U(1)$ scalar coset}\label{coset}

In this Appendix we collect some basic facts about the
$SU(1,1)/U(1)$ coset that will be needed in the main text.

The coset $SU(1,1)/U(1)$ consists of all $SU(1,1)$ matrices $V$
which are identified under the transformations of the compact
subgroup $U(1)$. If one takes $V$ to depend on space-time points
$x$ then the equivalence under $U(1)$ becomes a gauge symmetry. A
(left-)coset representative $V$ transforms as
\begin{equation}
    V(x)\rightarrow gV(x)h(x)\;,
\end{equation}
where $g\in SU(1,1)$ and $h\in U(1)$. We will parameterize the coset
representative $V$ as
\begin{equation}\label{matrixV}
     V=\left( \begin{array}{cc}
        V_{-}^{1} & V_{+}^{1} \\
        V_{-}^{2} & V_{+}^{2}
    \end{array} \right )\;,
\end{equation}
where $(\bar V_{\mp}^{1})=V_{\pm}^{2}$ and
$V_{-}^{\alpha}V_{+}^{\beta}-V_{-}^{\beta}V_{+}^{\alpha}=
\epsilon^{\alpha\beta}$ with the $SU(1,1)$ indices $\alpha,
\beta=1,2$.

Using the matrix $V$ a left-invariant Lie algebra element of
$SU(1,1)$ can be written as
\begin{equation}\label{liealgebra}
    V^{-1}\partial_{\mu}V=\left( \begin{array}{cc}
        -iQ_{\mu} & P_{\mu} \\
        \bar P_{\mu} & iQ_{\mu}
    \end{array} \right )\;,
\end{equation}
where $Q_{\mu}$ is real and transforms as a composite $U(1)$ gauge
field under local $U(1)$ transformations. The fields $P$ and $Q$ are
both invariant under global $SU(1,1)$ transformations. In terms of
the components of $V$ this equation reads
\begin{align}
        P_{\mu} & =-\epsilon_{\alpha\beta}V_{+}^{\alpha}\partial_{\mu}V_{+}^{\beta}\,,\label{defP} \\
        \bar P_{\mu} & =\epsilon_{\alpha\beta}V_{-}^{\alpha}\partial_{\mu}V_{-}^{\beta}\,,\label{defbarP} \\
        Q_{\mu} & =-i\epsilon_{\alpha\beta}V_{-}^{\alpha}\partial_{\mu}V_{+}^{\beta}\label{defQ}\,.
\end{align}

The gauge-covariant derivative of $P_\mu$ is defined in the standard
way as
$D_{\mu}P_{\nu}=\left(\partial_{\mu}-2iQ_{\mu}\right)P_{\nu}$. The Bianchi identity for $P_\mu$ is given by
\begin{equation}\label{Bianchi}
D_{[\mu}P_{\nu]}=0\;.
\end{equation}

The $U(1)$ gauge symmetry can be fixed by imposing the gauge
$V_{-}^{1}=V_{+}^{2}\in\mathbb{R}$. In this gauge the
 matrix elements of $V$ can
be parameterized by a complex scalar $\tau$ as
\begin{align}\label{parametrization}
    & V_{-}^{1}=V_{+}^{2}=\frac{\vert 1-i\tau\vert}{2(\rm{Im}\,\tau)^{1/2}}\,, \\
    & V_{+}^{1}=\bar V_{-}^{2}=\frac{1-i\bar\tau}{1+i\bar\tau}\frac{\vert 1-i\tau\vert}{2(\rm{Im}\,\tau)^{1/2}}\,.
\end{align}
Using this parametrization we have
\begin{align}
    & P_{\mu}=\frac{1}{\tau-\bar\tau}\frac{1-i\tau}{1+i\bar\tau}\partial_{\mu}\bar\tau\,,\label{Pintermsoftau} \\
    & Q_{\mu}=\frac{i}{2}\frac{1}{\tau-\bar\tau}\frac{1-i\bar\tau}{1-i\tau}\partial_{\mu}\tau+
    \frac{i}{2}\frac{1}{\tau-\bar\tau}\frac{1+i\tau}{1+i\bar\tau}\partial_{\mu}\bar\tau\,.
\end{align}

It is convenient to define the following gauge-invariant
(right-invariant) matrix $p_{\mu}$ as follows
\begin{equation}\label{defpp}
    p_{\mu}=V\mathcal{P}_{\mu}V^{-1}\;,\hskip 1truecm {\rm with}\hskip 1truecm
 \mathcal{P}_{\mu}=\left(\begin{array}{cc}
                            0 & P_{\mu} \\
                            \bar P_{\mu} & 0
                            \end{array}\right)\,.
\end{equation}
It can be shown that the components of $p_{\mu}$ are the three
Noether currents which are associated to the global $SU(1,1)$
invariance of the scalar kinetic terms of the IIB Lagrangian. In terms of the matrix $V$ the matrix
$\mathcal{P}_\mu$ is given by
\begin{equation}
 \mathcal{P}_{\mu}=\frac{1}{2}\left(V^{-1}\partial_{\mu}V+\left(V^{-1}\partial_{\mu}V\right)^{\dagger}\right)\,.
\end{equation}
In terms of $p_\mu$ the Bianchi identity \eqref{Bianchi} can be
written as
\begin{equation}\label{Bianchip}
\partial_{[\mu}p_{\nu]}-2p_{[\mu}p_{\nu]}=0\,.
\end{equation}

\setcounter{equation}0
\section{IIB supergravity}\label{sec:IIB}

\subsection{Manifest $SU(1,1)$ covariant formulation}

In the conventions of \cite{Schwarz:1983qr} and
\cite{Bergshoeff:2005ac} the bosonic part of the IIB supergravity action
\cite{Dall'Agata:1997ju,Dall'Agata:1998va} is given by
\begin{eqnarray}\label{IIBaction}
    S=\int_{{\mathcal{M}}_{10}}\left(*1R-2\bar P\wedge *P-\frac{1}{2}\bar G_3\wedge *G_3-4F_5\wedge *F_5+
\frac{i}{2}F_5\wedge\epsilon_{\alpha\beta}A_2^{\alpha}\wedge
F_3^{\beta}\right)\,\nonumber\\
-\int_{{\mathcal{M}}_{10}}d^{10}x\,{1\over{6\,\partial_ra\,\partial^ra}}
\partial^la(x)\,{\mathcal F}_{lm_1...m_4}{\mathcal
F}^{m_1...m_4p}\,\partial_pa(x)\,.
\end{eqnarray}
The forms $P$ (introduced in Appendix B), $G_3$ and $F_5$ are
defined via the following Bianchi identities
\begin{align}
    & DP=dP-2iQ\wedge P=0\,,\\
    & DG_3=dG_3-iQ\wedge G_3=-P\wedge\bar G_3 \label{BianchiG3}\,,\\
    & dF_5=-\frac{i}{8}G_3\wedge\bar G_3\,.\label{BianchiF5}
\end{align}
The solution to the Bianchi identity for $G_3$ is given by
$G_3=-\epsilon_{\alpha\beta}V_{+}^{\alpha}F^{\beta}_3$ where
$F_{3}^{\beta}=dA_2^{\beta}$. The 2-forms, $A_{2}^{\alpha}$, transform as a doublet under
$SU(1,1)$ and transform under gauge transformations as $\delta
A_{2}^{\alpha}=d\Lambda_1^{\alpha}$. The solution to the Bianchi
identity for $F_5$ reads
\begin{equation}
    F_5=dA_4+\frac{i}{16}\epsilon_{\alpha\beta}A_2^{\alpha}\wedge F_3^{\beta}\,.
\end{equation}
The last term in (\ref{IIBaction}) containing ${\mathcal F}_5\equiv
F_5-{}^*F_5$ and the auxiliary scalar field
$a(x)$ is the PST term. It ensures that on the mass shell the equations
of motion of $A_4$ reduce to the self-duality condition on its
field strength (see \cite{Dall'Agata:1997ju,Dall'Agata:1998va} for
details)
\begin{equation}
  {\mathcal F}_5= F_5-*F_5=0\,.
\end{equation}

\subsection{Introducing 6-forms}\label{sec:6forms}

It is possible to dualize the 2-forms, $A_2^{\alpha}$, to a doublet of 6-forms $A_6^{\alpha}$ via the duality relation and Bianchi identity
\begin{align}
    & F_7^{\alpha}=i*\left(V_{-}^{\alpha}G_3-V_{+}^{\alpha}\bar G_3\right) \label{dualityF7}\,,\\
    & dF_{7}^{\alpha}=4F_{3}^{\alpha}\wedge F_5\,.\label{BianchiF7}
\end{align}
We define
\begin{equation}\label{g7}
G_7=-\epsilon_{\alpha\beta}V_{+}^{\alpha}F_{7}^{\beta}
\end{equation}
which satisfies the following Bianchi identity
\begin{equation}\label{BianchiG7}
    DG_7+P\wedge\bar G_7=4G_3\wedge F_5\,.
\end{equation}
From equation \eqref{dualityF7} it follows that $G_7=i*G_3$.

We can write the NSD IIB action in the first order formalism as a
function of $A_2^{\alpha}$ and $F_7^{\alpha}$ such that the
variation with respect to $A_2^{\alpha}$ gives the Bianchi identity
\eqref{BianchiF7} and the variation with respect to $F_7^{\alpha}$
the duality relation \eqref{dualityF7}. This is achieved by the
following action
\begin{equation}
    S=\int_{{\mathcal{M}}_{10}}\left(*1R-2\bar P\wedge *P-\frac{1}{2}\bar G_7\wedge *G_7-4F_5\wedge *F_5+
\frac{i}{2}F_5\wedge\epsilon_{\alpha\beta}A_2^{\alpha}\wedge F_3^{\beta}+
\frac{i}{2}\epsilon_{\alpha\beta}A_2^{\alpha}\wedge dF_7^{\beta}\right)\,.\label{dualF7action}
\end{equation}
The 2-forms $A_2^{\alpha}$ are auxiliary variables. Their equations of motion, the Bianchi identities
\eqref{BianchiF7}, can be solved for $F_7^{\alpha}$ in terms of a doublet of 6-form potentials $A_6^{\alpha}$, via
\begin{equation}\label{F7}
    F_7^{\alpha}=dA^{\alpha}_6+\frac{4}{3}A_2^{\alpha}\wedge F_5-\frac{8}{3}F_3^{\alpha}\wedge A_4\,.
\end{equation}
One can substitute the on-shell value for $F_7^{\alpha}$ back into
the action \eqref{dualF7action} obtaining an action for
$A_6^{\alpha}$ in the second order formalism. If instead we
substitute the $F_7^{\alpha}$ equation of motion back into the
action we recover the action \eqref{IIBaction} (modulo the PST
part).

\subsection{Introducing 8-forms}\label{dualization}
We can introduce a triplet of 8-forms, $A^{\alpha\beta}_8$, via the
duality relation
\begin{align}
    & F_{9}^{\alpha\beta}=i*\left(V_{+}^{\alpha}V_{+}^{\beta}\bar P-V_{-}^{\alpha}V_{-}^{\beta}P\right)\,,
\label{F9duality}\\
    & dF_{9}^{\alpha\beta}=\frac{1}{4}F_3^{(\alpha}\wedge F_7^{\beta)}\,.\label{BianchiF9}
\end{align}
Solving the Bianchi identity \eqref{BianchiF9} we find that $F^{\alpha\beta}_9$ in the $SU(1,1)$ covariant formulation can be written as
\begin{equation}
    F^{\alpha\beta}_{9}=dA_8^{\alpha\beta}+\frac{1}{16}F_7^{(\alpha}\wedge A_2^{\beta)}-
\frac{3}{16}F_3^{(\alpha}\wedge A_6^{\beta)}\,.
\end{equation}
$F^{\alpha\beta}_{9}$ satisfies the following $SU(1,1)$ invariant
constraint \cite{Dall'Agata:1998va}
\begin{equation}
    \epsilon_{\alpha\gamma}\epsilon_{\beta\delta}V_{-}^{\alpha}V_{+}^{\beta}F^{\gamma\delta}_9=0\,.
    \label{definitionGtilde}
\end{equation}

The duality relation between the 8-forms and an axionic scalar
follows from the duality relation \eqref{BianchiF9} by contracting
the latter with the $SU(1,1)$ symmetric charge tensor
$q_{\alpha\beta}$ introduced in Section \ref{sec:Q7coupling} and
making use of equations \eqref{tautoT} to \eqref{defchi'}
\begin{equation}\label{axion8formduality1}
    \frac{\text{det}\,Q}{(\text{Im}\tau_0)^2}\vert(\tau-\tau_0)(\tau-\bar\tau_0)\vert^2 d\chi'=\left(T^2-4\text{det}\,Q\right)d\chi'=*q_{\alpha\beta}F^{\alpha\beta}_9\,.
\end{equation}

Eq. \eqref{axion8formduality1} makes manifest the statement
that each 8-form is dual to an axionic scalar and that implementing
this duality relation into the IIB action makes it mandatory to
perform the field redefinition as given in Eq. \eqref{tautoT}.

Modulo the PST term, the action for the bosonic sector of IIB
supergravity which reproduces the duality relations between
$F^\alpha_2$ and
$F_7^{\alpha}$ and between $d\chi'$ and
$q_{\alpha\beta}F_9^{\alpha\beta}$ has the form
\begin{align}
&
S=\int_{{\mathcal{M}}_{10}}\left[*1R-\frac{1}{2}\frac{1}{T^2-4\det\,Q}\left(dT\wedge
*dT+
q_{\alpha\beta}F_9^{\alpha\beta}\wedge *q_{\alpha\beta}F_9^{\alpha\beta}\right)-\frac{1}{2}\bar G_7\wedge *G_7\nonumber\right.\\
&\left.-4F_5\wedge
*F_5+\frac{i}{2}F_5\wedge\epsilon_{\alpha\beta}A_2^{\alpha}\wedge
F_3^{\beta}+\frac{i}{2}\epsilon_{\alpha\beta}A_2^{\alpha}\wedge
dF_7^{\beta}+\chi' d\left(q_{\alpha\beta}
F_9^{\alpha\beta}\right)\right]\,,\label{dualF9action}
\end{align}
where $F^{\alpha\beta}_9$ and $F^\alpha_7$ are considered as
independent fields and
$G_7$ depends on $F^\alpha_7$ and $\chi'$ as given in Eq. \eqref{g7} in
which $\phi$ and $\chi$ are expressed in terms of $T$ and $\chi'$
using \eqref{tautoT} and \eqref{mathcalT}.

The form of \eqref{dualF9action} differs from the action
\eqref{ad+7+10}. The former contains $F^{\alpha\beta}_9$ as an
independent field, while the latter depends on $A^{\alpha\beta}_8$
and thus is more suitable for describing the minimal electric
coupling of the Q7-brane.

\subsection{Gauge field transformations}
In the
$SU(1,1)$ covariant formulation the r-form gauge fields ($r=2,4,6,8$)
have the following gauge transformations:
\begin{align}
    & \delta A_2^{\alpha}=d\Lambda_1^{\alpha}\,,\label{A2gauge}\\
    & \delta A_4=d\Lambda_3-\frac{i}{16}\epsilon_{\gamma\delta}\delta A_2^{\gamma}\wedge A_2^{\delta}\,,\label{A4gauge}\\
    & \delta A_6^{\alpha}=d\Lambda_5^{\alpha}+\frac{8}{3}A_2^{\alpha}\wedge\delta A_4-\frac{4}{3}A_4\wedge\delta A_2^{\alpha}+
    \frac{i}{12}A_2^{\alpha}\wedge\epsilon_{\gamma\delta}\delta A_2^{\gamma}\wedge A_2^{\delta}\,,\label{A6gauge}\\
    & \delta A_8^{\alpha\beta}=d\Lambda_7^{\alpha\beta}+\frac{3}{16}A_2^{(\alpha}\wedge\delta A_6^{\beta)}-
    \frac{1}{16}A_6^{(\alpha}\wedge\delta A_2^{\beta)}-\frac{1}{4}A_2^{\alpha}\wedge A_2^{\beta}\wedge\delta A_4+\nonumber\\
    & \qquad\quad\;\;\frac{1}{6}A_4\wedge A_2^{(\alpha}\wedge\delta A_2^{\beta)}-\frac{1}{12}A_2^{\alpha}\wedge A_2^{\beta}\wedge
    \frac{i}{16}\epsilon_{\gamma\delta}\delta A_2^{\gamma}\wedge A_2^{\delta}\label{8formgauge}\,.
\end{align}

\subsection{Manifest $SL(2,\mathbb{R})$ covariant formulation}\label{sec:SL2Rcov}

We now formulate the IIB theory in the standard $(\tau,\bar\tau)$ basis. The RR and NSNS 2-forms are denoted by $C_2$ and $B_2$, respectively. Their duals will be denoted by $C_6$ and $B_6$. They are defined by $B_{2,6}=\frac{1}{2}(A_{2,6}^1+A_{2,6}^2)$ and $C_{2,6}=\frac{i}{2}(A_{2,6}^1-A_{2,6}^2)$. The axion-dilaton field $\tau$ is $\tau=\chi+ie^{-\phi}$. The objects $P$, $Q$, $G_3$, $F_5$ and $G_7$ can be written
as\footnote{To obtain these expressions one can start from the parametrization of $V^{\alpha}_{\pm}$ given in
\eqref{parametrization} and the definitions of $P$, $Q$ and $G_3$. Then use the freedom to perform the local $U(1)$ transformation $P\rightarrow e^{2i\alpha}P$, $Q\rightarrow Q+d\alpha$ and $G_{3,7}\rightarrow e^{i\alpha}G_{3,7}$ with parameter $\alpha$ given by $e^{2i\alpha}=\frac{1+i\bar\tau}{1-i\tau}$.}
\begin{align}
    & P=\frac{d\bar\tau}{\tau-\bar\tau}\,,\label{Pstandardgauge}\\
    & Q=i\frac{d(\tau+\bar\tau)}{2(\tau-\bar\tau)}\,,\\
    & G_3=\frac{i}{(\text{Im}\tau)^{1/2}}\left(-dC_2+\bar\tau
    dB_2\right)\,,\label{G3standardgauge}\\
    & F_5=dA_4+\frac{1}{8}\left(C_2\wedge dB_2-B_2\wedge dC_2\right)\label{F5}\,,\\
    & G_7=\frac{i}{(\text{Im}\tau)^{1/2}}\left(-dC_6+\bar\tau
    dB_6+\frac{4}{3}(-C_2+\bar\tau B_2)\wedge F_5\nonumber\right.\\
&\left.-\frac{8}{3}(-dC_2+\bar\tau dB_2)\wedge A_4\right)\,.\label{G7standardgauge}
\end{align}

In this formulation the following symmetry is manifest
\begin{equation}\label{sltrafos}
    \tau\rightarrow\frac{a\tau+b}{c\tau+d}\quad\text{and}\quad
    \left(\begin{array}{c}
    C_{2} \\
    B_{2}
    \end{array}\right)\rightarrow\left(\begin{array}{cc}
    a & b \\
    c & d
    \end{array}\right)\left(\begin{array}{c}
    C_{2} \\
    B_{2}
    \end{array}\right)\quad\text{with}\quad ad-bc=1\,.
\end{equation}

\end{document}